\documentclass[12pt]{article}
\usepackage{hyperref}
\usepackage{amssymb,latexsym,amsmath} 
\usepackage{authblk}
\usepackage[utf8]{inputenc}
\usepackage{dsfont}
\usepackage[utf8]{inputenc}
\usepackage{graphics}
\usepackage{color}
\usepackage{subfig}
\usepackage{graphicx}
\usepackage{mathrsfs}
\begin{document}
\title{\bf Complexity of Einstein-Maxwell-non-minimal coupling $R^2F^2$: the role of the penalty factor}

\author{Mojtaba Shahbazi \thanks{Corresponding author: mojtaba.shahbazi@modares.ac.ir},  Mehdi Sadeghi\thanks{mehdi.sadeghi@abru.ac.ir}\,\,\, \hspace{2mm}\\
	{\small {\em Department of Physics, Faculty of Basic Sciences,}}\\
	{\small {\em Ayatollah Boroujerdi University, Boroujerd, Iran}}
}
\date{\today}
\maketitle

\abstract{We investigate holographic complexity in Einstein-Maxwell theory with a non-minimal coupling of the form $R^2F_{\mu\nu}F^{\mu\nu}$ within the complexity=anything framework. A perturbative AdS black brane solution is constructed to first order in the non-minimal coupling parameter. Owing to the linear temperature dependence of the resistivity, this model provides a holographic realization of strange metal behavior. The complexity growth rate (CGR) is governed by three independent parameters: the conserved charge, the non-minimal coupling, and the choice of the generalized term entering the complexity functional. We consider three representative generalizations, namely the Weyl tensor squared, $R^2F^2$
, and $F^2$. We provide a physical interpretation of these parameters, the generalized bulk functional analytically induces a deformation of the effective cost metric, which can be interpreted as a bulk penalty factor, while the conserved charge and the non-minimal coupling control an effective scrambling time in the dual theory. The role of the generalization parameter is shown to be closely tied to the structure of the corresponding quantum circuit.}\\


\section{Introduction} \label{intro}
The AdS/CFT (Anti-de Sitter/Conformal Field Theory) duality provides a powerful framework for understanding the relationship between gravitational theories in asymptotically AdS spaces and their corresponding quantum field theories on the boundary \cite{malda}. Although, our universe is a dS space, studying the duality in AdS is valuable to find a deep insights to develop the duality to dS/CFT and real-world physics \cite{stro, hcos, hcos1,hcos2,hcos3,hcos4,hcos5,hcos6}. This duality suggests that complex gravitational phenomena can be analyzed through the lens of quantum information theory, enabling the exploration of measures such as the quantum computational complexity. The complexity can be viewed as a measure of how difficult it is to prepare a given quantum state from a reference state using a sequence of operations \cite{3lec}. 
The complexity serves as a crucial indicator of the degrees of freedom \cite{deg1, deg2} and the scrambling dynamics in a quantum system, revealing insights into how information is processed and stored in black hole backgrounds. By examining complexity within the AdS/CFT framework, we gain valuable insights into the structure of quantum states and their holographic properties, ultimately bridging the gap between deep gravitational insights and operational aspects of quantum field theories.

Recent developments in theoretical physics have established intriguing relationships between quantum complexity and fundamental physical quantities, notably through the lens of the Einstein-Rosen bridge volume. 
It suggests that the growth of the Einstein-Rosen Bridge (ERB) in an eternal AdS black hole can be explained by ER=EPR. It is given by the maximum volume of a hypersurface (ERB) anchored in the boundary time slice $\Sigma_{CFT}$, where the boundary CFT state live \cite{cv}:
\begin{align}\label{cv}
	\mathcal{C}(\Sigma_{CFT})= \max_{\partial\Sigma=\Sigma_{CFT}} \frac{V(\partial\Sigma)}{G_N L},
\end{align}
here $\Sigma$ is extremal hypersurface anchored in the boundary time slice $\Sigma_{CFT}$. On the other hand, the other conjecture that relates complexity to the action of the gravitational theory restricted to the Wheeler-DeWitt (WDW) patch, expressed as \cite{ca}:
\begin{align}\label{ca}
	\mathcal{C}(\Sigma)= \frac{I_{WDW}}{\pi\hbar}.
\end{align} 
In \cite{cv2}, it is defined the spacetime volume for the WDW patch as:
\begin{align}
	\mathcal{C}(\Sigma)=\frac{V_{WDW}}{G_N L^2}.
\end{align}

More recently, these ideas have been unified and extended through the complexity=anything conjecture \cite{general}, which allows complexity to be defined via a broad class of diffeomorphism-invariant bulk functionals. This generalization introduces an intrinsic ambiguity in the definition of complexity, encoded in a choice of scalar functional that weights different regions of spacetime.

The exploration of non-minimal theories, those that connect the gravitational field with other fields through cross terms involving the curvature tensor, has gained traction as alternative gravity theories over time, particularly the non-minimal coupling has a significance on high curvature spacetimes. Non-minimal field theories can be categorized into five classes, based on the types of fields that non-minimally interact with gravitation:

1) The first category involves the coupling of scalar fields to spacetime curvature, originating from the Scherrer-Jordan-Thiry-Brans-Dicke theory, where a scalar field interacts non-minimally with the Ricci scalar. Various authors \cite{scherrer,jordan,thiry,bd} had distinct motivations for proposing this theory; for a historical review, see \cite{goenner}. Additionally, a scalar field conformally coupled to gravity was introduced in \cite{ccj}, with a review provided in \cite{NMscal1}.

2) The second category focuses on modeling non-minimal interactions between the electromagnetic field and curvature, often referred to as the non-minimal Einstein-Maxwell model (see, for example, \cite{NM1,NM2,NM3,BL2005}). 

3) The third class includes Einstein-Yang-Mills models characterized by SU(N) symmetry \cite{Horn81,MH2}. 

4) The fourth category consists of Einstein-Yang-Mills-Higgs models \cite{BaDeZa07,BaDeZa08}. 

5) Lastly, the fifth class encompasses models featuring an axion pseudo-scalar field, which couples to both electromagnetic and gravitational fields, known as non-minimal Einstein-Maxwell-axion models \cite{BNi2010}.

Non-minimal coupling in gravity, particularly when invoking the AdS (Anti-de Sitter) framework, has gained significant attention due to its potential to enhance our understanding of gravitational dynamics in various physical contexts. Due to the curvature coupling, these non-minimal theories have a significant effect on the higher curvature regimes such as near black hole or the early universe. The non-minimal term could be viewed as deforming the boundary theory, which means that n-point functions and transport coefficients would be altered. Specifically, this term breaks the translational symmetry which is more appropriate to model realistic crystals where containing impurity \cite{superimpurity}. In addition, the non-minimal theories could model strange metals where the resistivity scales linearly with temperature \cite{strangmetal}. The model we have considered in this paper represents a linear relation between the resistivity and the temperature, a hallmark of strange metals \cite{linresis}. Motivated by this observation, we study a four-dimensional Einstein-Maxwell theory with a non-minimal coupling of the form $R^2F_{\mu\nu}F^{\mu\nu}$. Since exact solutions are not available, we construct a perturbative AdS black brane solution to first order in the non-minimal coupling.

The main goal of this work is to explore how such non-minimal interactions and the choices of the generalized parameter affect holographic complexity. Within the complexity=anything framework, we analyze the time dependence of the complexity growth rate (CGR) and show that it is controlled by three parameters: the conserved charge, the non-minimal coupling, and the choice of generalization in the complexity functional. We consider three representative choices, the Weyl tensor squared, $R^2F^2$, $F^2$, and provide a physical interpretation of their effects.

An important aspect of quantum circuit complexity is the freedom in choosing the set of elementary gates and the associated cost or penalty factors that weight different directions in the space of unitary transformations. In Nielsen's geometric formulation, this freedom is encoded in the choice of a cost metric, and different penalty schemes correspond to inequivalent notions of complexity. In holographic settings, however, the microscopic identification of boundary gates remains intrinsically ambiguous \cite{general}. As a result, holographic complexity proposals should be understood as defining equivalence classes of circuit complexities rather than unique microscopic constructions. This intrinsic ambiguity motivates the study of generalized complexity functionals and their geometric implications.

The three generalization choices allow us to systematically explore how different generalizations modify the effective geometry governing extremal bulk trajectories associated with complexity. We show analytically that the generalized functional deforms the effective cost metric, in a manner structurally equivalent to the introduction of penalty factors in Nielsen’s circuit complexity. This identification does not rely on specifying a unique set of boundary gates; rather, it emerges at the level of complexity geometry and reflects the inherent freedom in defining cost structures in holographic complexity.

We further examine the implications of this deformation for the time dependence of complexity growth. In particular, we analyze the CGR and identify an effective scrambling time that controls this behavior. We emphasize that this timescale should be understood as a complexity-dependent effective scrambling time, rather than the microscopic scrambling time defined via out-of-time-order correlators or shock-wave analyses. While the intrinsic chaotic properties of the bulk geometry remain unchanged, the effective scrambling time governing complexity growth is shown analytically to depend on both the non-minimal coupling and the generalized parameter through their influence on the effective complexity geometry.

The paper is organized as follows: in the section \ref{sec2} we construct the perturbative black brane solution. In the section \ref{sec3}, we introduce the generalized complexity functional and analyze the resulting complexity growth for the three choices of the generalized parameter. The section \ref{sec4} is devoted to the interpretation of the generalized functional as encoding a penalty structure and to the analysis of the effective scrambling time governing complexity growth and supporting this interpretations by presenting qualitative analogies with superconducting quantum circuits. We conclude in the section \ref{conclusion} with a discussion of our results and possible future directions.
\section{Non-minimal $R^2F_{\mu \nu}F^{\mu \nu} $  AdS Black Brane Solution}
\label{sec2}
\indent The non-minimal Einstein-Maxwell theory, featuring a negative cosmological constant, is represented by the following action \cite{Balakin:2015gpq},\cite{Sert:2020vmq},\cite{Lambiase:2008zz}:
\begin{eqnarray}\label{action}
S=\int d^{4}  x\sqrt{-g} \bigg[\frac{1}{\kappa }(R-2\Lambda )-\frac{q_1}{4}F_{\mu \nu }F^{\mu \nu} +q_2 R^2 F_{\mu \nu }F^{\mu \nu} \bigg],
\end{eqnarray}
where $\kappa$ is the gravitational constant, $R$ is the Ricci scalar, $\Lambda=-\frac{3}{L^2}$ is the cosmological constant, $L$ representing the AdS radius. The parameter  $q_1$ is dimensionless.\\
The Maxwell field strength $ F_{\mu \nu }$ is given by:
\begin{align} \label{YM}
F_{\mu \nu } =\partial _{\mu } A_{\nu } -\partial _{\nu } A_{\mu },
\end{align}
with the gauge coupling constant set to 1 and $A_{\nu }$ representing the Maxwell potentials. The parameter $q_2$ is a dimensionful coupling constant which also acts as the interaction term between the gauge field and the Ricci scalar.\\
Varying the action (\ref{action}) with respect to the spacetime metric  $g_{\mu \nu } $ yields the field equations:
\begin{equation}\label{EOM1}
R_{\mu \nu }-  \tfrac{1}{2} g_{\mu \nu } R + \Lambda g_{\mu \nu }=\kappa T^{\text{(eff)}}_{\mu \nu }
\end{equation}
where,
\begin{equation}
T^{\text{(eff)}}_{\mu \nu }=q_1T^{\text{(M)}}_{\mu \nu } + q_2T^{(I)}_{\mu \nu }
\end{equation}
The energy-momentum tensor for the Maxwell field is given by,
\begin{equation}
T^{\text{(M)}}_{\mu \nu }=  \tfrac{1}{4}g_{\mu \nu } F_{\alpha \beta } F^{\alpha \beta }-F_{\mu }{}^{\alpha } F_{\nu \alpha },
\end{equation}
and the additional interaction term is,
\begin{eqnarray}
	&&T^{(I)}_{\mu \nu }=2 F_{\alpha \beta } F^{\alpha \beta} R_{\mu \nu } R + 2 F_{\mu }^{\alpha} F_{\nu \alpha } R^2 -  \tfrac{1}{2} F_{\alpha \beta } F^{\alpha \beta}  g_{\mu \nu } R^2 \nonumber \\ 
	&& + 8 F^{\beta \gamma} g_{\mu \nu } \nabla_{\alpha }F_{\beta \gamma } \nabla^{\alpha }R + 4 F^{\alpha \beta} g_{\mu \nu } R \nabla_{\gamma }\nabla^{\gamma }F_{\alpha \beta } \nonumber \\ 
	&& + 2 F_{\alpha \beta } F^{\alpha \beta} g_{\mu \nu } \nabla_{\gamma }\nabla^{\gamma }R + 4 g_{\mu \nu } R 
	\nabla_{\gamma }F^{\alpha \beta} \nabla^{\gamma }F_{\alpha \beta } \nonumber \\ 
	&& - 2 F^{\alpha \beta} R \nabla_{\mu }\nabla_{\nu }F_{\alpha \beta } - 4 R \nabla_{\mu }F^{\alpha \beta} \nabla_{\nu }F_{\alpha \beta } - 4 F^{\alpha \beta} \nabla_{\mu }R \nabla_{\nu }F_{\alpha \beta } \nonumber \\ 
	&& - 4 F^{\alpha \beta} \nabla_{\mu }F_{\alpha \beta } \nabla_{\nu \
	}R - 2 F^{\alpha \beta} R \nabla_{\nu }\nabla_{\mu }F_{\alpha \beta } - 2 F_{\alpha \beta } F^{\alpha \beta}\nabla_{\nu }\nabla_{\mu }R.
\end{eqnarray}
Varying the action (\ref{action}) with respect to $A_{\mu}$ results in the field equations:
 \begin{eqnarray}\label{EOM-Maxwell}
\nabla_{\mu }\Big(-\frac{q_1}{2} F^{\mu \nu } + 2 q_2 F^{\mu \nu } R^2\Big)=0.
\end{eqnarray}
Considering the equations of motion for the field $A^{\mu}$ reveals that the sign of the coupling constant $q_2$ should be negative. The coefficient of the kinetic term would be:
\begin{align}
-\frac{1}{4}\Big(q_1-4q_2 R^2\Big),
\end{align}
at large curvature regime, when $4q_2R^2 \gg q_1$ the coefficient of the kinetic term in the Maxwell equation flips sign which means the ghost instability (i.e., the spectrum is not bounded
from below). Then, the sign of the coupling constant $q_2$ should be negative.
Considering a four dimensional spacetime with a two-dimensional spatial subspace, we adopt the following ansatz for the metric:
\begin{equation}\label{metric}
ds^{2} =-e^{-2H(r)}f(r)dt^{2} +\frac{dr^{2} }{f(r)} +\frac{r^2}{L^2}(dx^2+dy^2).
\end{equation}
We will use this ansatz to solve Eq. (\ref{EOM-Maxwell}), where the potential 1-form is expressed as \footnote{We have chosen the radial gauge
$A_r=0$ and assumed translational symmetry in the boundary directions,
which together with staticity leads to the vanishing of the
spatial components. Furthermore, to ensure the regularity of the
gauge potential at the future horizon, we have imposed the condition
that $h(r)$ vanishes at the horizon $r=r_h$.}:
\begin{equation}\label{background}
A_{\mu} = \left(h(r), 0, 0, 0\right).
\end{equation}
Now, we can formulate the equations of motion based on Eqs. (\ref{EOM-Maxwell}) - (\ref{background}) (see Appendix \ref{app1}).\\

The solution of $h(r)$ is as \footnote{The solution is given by the solving \eqref{EOM-Maxwell} by use of ansatz \eqref{metric} and \eqref{background}.}:
\begin{equation}
h(r)= C_1\int^{r}\frac{e^{-H(u)}u^2}{B(u)}du+C_2,
\end{equation}
\begin{equation}
B(u)=4 q_2 \left(u \left(f' \left(4-3 u H'\right)+u f''\right)+2 f \left(\left(u H'-1\right)^2-u^2
H''\right)\right)^2-q_1 u^4
\end{equation}
We will solve the metric and the gauge field equations perturbatively up to first order in $q_2$ and consider the following forms for $f (r)$, $H(r)$ and $h (r)$:
\begin{equation}\label{f}
f(r)=f_0(r)+q_2 f_1(r),
\end{equation}
\begin{equation}\label{h}
h(r)=h_0(r)+q_2 h_1(r),
\end{equation}
\begin{equation}\label{H}
H(r)=H_0(r)+q_2 H_1(r),
\end{equation}
where $f_0(r)$, $h_0(r)$ and $H_0(r)$ are the leading order solutions of Einstein-Maxwell AdS black brane in four dimensions.\\
The $h_0(r)$ , $f_0(r)$ and $H_0(r)$ are found exactly as,
\begin{equation}
h_0(r)=C_2-C_1\int^r\frac{ 1}{q_1 u^2}du=Q(\frac{1}{r}-\frac{1}{r_h}),
\end{equation}
where $C_1=q_1 Q$ and $C_1=-C_2$,
\begin{equation}\label{f0}
f_0(r)=-\frac{2m_0}{r}-\frac{\Lambda r^2}{3}-\frac{\kappa q_1}{r}\int^r  u^2 
h_0'^2 du=-\frac{2m_0}{r}+\frac{r^2}{L^2}+\frac{\kappa q_1 Q^2}{r^2},
\end{equation}
Since the equations for the $tt$--component and the $rr$-component are identical up to first order, we conclude that 
$8 f_0(r) r^3 H_0'(r)=0$. Therefore, we determine that 
$H_0$ is given as follows \footnote{By the condition
$8f_0(r)r^3H'_0(r)=0$ we deduce that $H_0(r)$ is a constant. We apply this condition that the speed of light on the boundary is equal to one.
Therefore, $H_0(r)=0$.}:
\begin{equation}
H_0(r)=0.
\end{equation}
The blackening factor at the event horizon must be suppressed, i.e., $f(r_h)=0$. The mass $M$ of the black brane can be determined by applying this condition:
\begin{equation}\label{m}
m_0=\frac{\kappa q_1 Q^2}{2 r_h}+\frac{r_h^3}{2L^2}.
\end{equation}
By plugging Eq.(\ref{m})  in Eq.(\ref{f0}) we have:
\begin{equation}
f_0(r)=\frac{r^2}{L^2}(1-\frac{r_h^3}{r^3})+\frac{\kappa q_1 Q^2}{4 r}(\frac{1}{r}-\frac{1}{r_h}).
\end{equation}
By subtraction of \ref{tt-comp} from \ref{rr-comp} and solving it, $H_1$ is calculated as follows:
\begin{eqnarray}
	&&H_1(r)=-12 \mathit{k} f_0''(r) h_0'(r)^2-32 \mathit{k} f_0'(r) h_0'(r) h_0''(r)-8 \mathit{k}
	r f_0''(r) h_0'(r) h_0''(r)\nonumber\\&&+\frac{24 \mathit{k} f_0'(r) h_0'(r)^2}{r}-4 \mathit{k} r
	f_0^{(3)}(r) h_0'(r)^2-\frac{16 \mathit{k} f_0(r) h_0'(r) h_0''(r)}{r}\nonumber\\&&+\frac{24 \mathit{k}
		f_0(r) h_0'(r)^2}{r^2}.
\end{eqnarray}
Hence, $H_1$ is calculated as follows:
\begin{equation}
	H_1(r)=-\frac{80 \kappa  Q^2 \Lambda}{r^4}=\frac{240 \kappa  Q^2 }{L^2 r^4}.
\end{equation}
$h_1(r)$ can be calculated as follows,
\begin{eqnarray}
h_1(r)&&= C_1 \int^r_{r_h} \left(\frac{H_1(u)}{q_1 u^2}-\frac{4 \left(u \left(4 f_0'(u)+u f_0''(u)\right)+2
	f_0(u)\right)^2}{q_1^2 u^6}\right)du\nonumber\\&&=\frac{576 C_1}{L^4 q_1^2 }(\frac{1}{r}-\frac{1}{r_h})-\frac{24 \mathit{k} Q^2 C_1 }{L^2
	q_1}(\frac{1}{r^5}-\frac{1}{r_h^5}),
\end{eqnarray}
where $C_1=q_1 Q$.
\begin{equation}
h_1(r)=\frac{576 Q}{L^4 q_1 }(\frac{1}{r}-\frac{1}{r_h})-\frac{48 \mathit{k} Q^3  }{L^2 }(\frac{1}{r^5}-\frac{1}{r_h^5}),
\end{equation}
$f_1(r)$ can be calculated as:

\begin{eqnarray}
&&f_1(r)=-\frac{672 \mathit{k} m_0 Q^2}{L^2 r^5}+\frac{288 \mathit{k}^2 Q^4 q_1}{L^2 r^6}+\frac{768 \mathit{k}
	Q^2}{L^4 r^2}-\frac{2 m_1}{r}.
\end{eqnarray}

We have obtained the solution up to the linear order of $q_2$:
\begin{eqnarray}
	&&h(r)=Q(\frac{1}{r}-\frac{1}{r_h})+q_2 \Big\{\frac{576 Q}{L^4 q_1 }(\frac{1}{r}-\frac{1}{r_h})-\frac{48 \mathit{k} Q^3  }{L^2 }(\frac{1}{r^5}-\frac{1}{r_h^5})\Big\},\label{bulkf}\\
		&&H(r)=-q_2\frac{80 \kappa  Q^2 \Lambda}{r^4},\label{H}\\
			&&f(r)=-\frac{2m_0}{r}+\frac{r^2}{L^2}+\frac{\kappa q_1 Q^2}{r^2}\nonumber\\&&+q_2\Big(-\frac{672 \mathit{k} m_0 Q^2}{L^2 r^5}+\frac{288 \mathit{k}^2 Q^4 q_1}{L^2 r^6}+\frac{768 \mathit{k}
				Q^2}{L^4 r^2}-\frac{2 m_1}{r}\Big)\label{fq}.
\end{eqnarray}
We reparametrize such that:
\begin{align}
m_0+q_2 m_1=:\bar{m}_0,
\end{align}
So $f(r)$ changes as:
\begin{eqnarray}
&&f(r)=-\frac{2\bar{m}_0}{r}+\frac{r^2}{L^2}+\frac{\kappa q_1 Q^2}{r^2}\nonumber\\&&+q_2\Big(-\frac{672 \mathit{k} m_0 Q^2}{L^2 r^5}+\frac{288 \mathit{k}^2 Q^4 q_1}{L^2 r^6}+\frac{768 \mathit{k}
	Q^2}{L^4 r^2}\Big).
	\end{eqnarray}
This is the solution up to the linear order of $q_2$.

\section{Holographic Complexity of $R^2F^2$}
\label{sec3}

$Complexity=anything$ conjecture has emerged as a generalized version of the complexity in the bulk. It states that the complexity can be described by a more general functional to include a class of new diffeomorphism invariant observables \cite{general}:
\begin{align}\label{obser}
	\mathcal{O}_{F_1,\Sigma_{F_2}}(\Sigma_{CFT})=\frac{1}{G_N L}\int_{\Sigma_{F_2}}\mathrm{d}^{d}\sigma \sqrt{h} F_1(g_{\mu\nu};X^{\mu})
\end{align}
where, $F_1$ can be a general scalar function of metric $g_{\mu\nu}$  and an embedding $X^{\mu}(\sigma^a)$ of the hypersurfaces. Moreover, $\Sigma_{F_2}$ could be a general hypersurface but we restrict ourself to codimension-one hypersurfaces in the bulk spacetime with boundary time slice $\partial\Sigma_{F_2} =\Sigma_{CFT}$. By extremizing the hypersurface it leads to:
\begin{align}\label{var}
	\delta_{X}\Big(\int_{\Sigma}\mathrm{d}^{d}\sigma \sqrt{h} F_2(g_{\mu\nu};X^{\mu})\Big)=0. 
\end{align}
For simplicity, we follow the case $F_1 =F_2$, so the observable \eqref{obser}, known by $\mathcal{C}_{Any}$, obeying the above condition is given by:
\begin{align}\label{maxv}
	\mathcal{C}_{Any}(\tau)= \max_{\partial\Sigma(\tau)=\Sigma_{\tau}}\frac{V_x}{G_N L}\left[\int_{\Sigma}\mathrm{d}^{d}\sigma \sqrt{h} F_1(g_{\mu\nu};X^{\mu})\right]
\end{align}
where $h$ is the determinant of induced metric on the given hypersurface.
The usual choice of the generalization is:
\begin{align}\label{F1F2}
	F_1=F_2=1+\gamma b,
\end{align}
where $b$ could be $C^2$ the Weyl tensor squared, $F^2$, the field strength and $R^2F^2$, where $R$ is the Ricci scalar. For $\gamma=0$ the CV is retrieved. In the following, we are going to study the $\mathcal{C}_{Any}$ of this setup, with the metric solution \eqref{metric}. To do this, we need to transform the metric to Eddington-Finkelstein coordinates:
\begin{align}\label{metricEF}
	\mathrm{d} s^{2}=-e^{-2H}f(r) \mathrm{d} v^{2}+ 2e^{-H} \mathrm{~d} v \mathrm{~d} r+ \frac{r^2}{L^2}\big( dx^2+dy^2\big),
\end{align}
where:
\begin{align}
	v=t+r^{*}(r), \quad \mathrm{d}r^{*}=\frac{\mathrm{d}r}{f(r)e^{-H}}.
\end{align}
Then, the complexity could be computed from \eqref{maxv} and \eqref{F1F2} as \cite{general}:
\begin{align}\label{Any}
	\mathcal{C}_{Any}(\tau)= \frac{V_0}{G_N L}\int_{\Sigma}\mathrm{d}\sigma r^{d-1}\sqrt{-e^{-2H}f \dot{v}^{2}+2 e^{-H}\dot{v} \dot{r}} a(r),
\end{align}
where the dots indicate derivatives with respect to $\sigma$, and $V_0$ represents the volume of spatial $x$ and $y$ directions.

Supposing $\mathcal{C}_{Any}$ as an action, then the conserved  momentum conjugate to coordinate $v$ would be:
\begin{align}
	P_v=-\frac{\partial \mathcal{L}}{\partial \dot{v}}=\frac{a(r) r^{d-1}(-e^{-2H}f \dot{v}+ e^{-H}\dot{r})}{\sqrt{-e^{-2H}f \dot{v}^{2}+2 e^{-H}\dot{v} \dot{r}}}=\dot{r}-e^{-H}f(r) \dot{v}.	
\end{align}
Due to the fact that \eqref{Any} is diffeomorphism invariant and doesn't change under reparametrization, the parameter $\sigma$ could be fixed by choosing: 
\begin{align}
	\sqrt{-e^{-2H}f \dot{v}^{2}+2 e^{-H}\dot{v} \dot{r}}=a(r) r^{d-1}e^{-H}.
\end{align}
Then, extremality conditions is given:
\begin{align}\label{rdot}
	\dot{r}=\pm \sqrt{P^2_v +e^{-2H}f(r)a(r)^2 r^{2(d-1)}},
\end{align}
\begin{align}\label{tau}
	\dot{t}=\dot{v}-\frac{\dot{r}}{e^{-H}f(r)}=\frac{-P_v\dot{r}}{e^{-H}f(r) \sqrt{	P^2_v- \tilde{U}(r)}}.
\end{align}
where $\tilde{U}(r)$ is an effective potential:
\begin{align}\label{poten}
	\tilde{U}(r)=-e^{-2H}f(r)a(r)^2r^{2(d-1)}.
\end{align}
Then, \eqref{rdot} gives the equation of motion like for a classical particle:
\begin{align}
	\dot{r}^{2}+\tilde{U}(r)=P^2_v.
\end{align}
On the symmetric trajectory, the conserved momentum is a function of the turning point $r_{min}$:
	\begin{align}
		P^2_v=	\tilde{U}(r_{min})=-e^{-2H(r_{min})}f(r_{min})a(r_{min})^2 r_{min}^{2(d-1)}.
	\end{align}	
The boundary time is related to the conserved momentum and the turning point $r_{min}$ acts as an intermediary:
	\begin{align}\label{tauint}
		\tau & =2 \int_{r_{\min }}^{\infty} \mathrm{d} r \frac{-P_v}{e^{-H}f(r) \sqrt{	P^2_v -\tilde{U}(r)}}.
	\end{align}
Boundary time derivation of $\mathcal{C}_{Any}$ leads to:
	\begin{align}
		\frac{\mathrm{d} \mathcal{C}_{Any}}{\mathrm{~d} \tau}=\frac{1}{2} \frac{\mathrm{d} \mathcal{C}_{Any}}{\mathrm{~d} \tau_{\mathrm{R}}}&=\frac{V_0}{G_N L} P_v\nonumber\\
		&=\frac{V_0}{G_NL} \sqrt{-e^{-2H(r_{min})}f(r_{min})}a(r_{min}) r_{min}^{(d-1)}.\label{cdotf}
	\end{align}
The time evolution of the generalized complexity growth rate is governed
by the behavior of the conserved momentum $P_v$. The late-time behavior of the growth rate is given by: 
	\begin{align}\label{tauinf}
		\lim _{\tau \rightarrow \infty} \frac{\mathrm{d} \mathcal{C}_{Any}}{\mathrm{~d} \tau}=\frac{V_0}{G_NL} \sqrt{-e^{-2H(\tilde{r}_{min})}f\left(\tilde{r}_{\min }\right)} a(\tilde{r}_{\min }) \tilde{r}_{min}^{(d-1)},
	\end{align}
where $\tilde{r}_{\min }$ is local maximum of the effective potential, and owing to the fact that at this radius time goes to infinity, it is known as $r_f:= \tilde{r}_{\min }$.\\

We now explain that the numerical analysis shows the relationship
between the boundary time $\tau$ and the conserved momentum $P_v$
(which sets the complexity growth rate, $\frac{d\mathcal{C}_{Any}}{d\tau}\propto P_v$) for the
cases $b$ equals to $C^2$, $R^2F^2$ and $F^2$. The panels demonstrate how this relationship
changes with different parameters ($Q$, $q_2$ 
 and $\gamma$). Regarding the fact that the complexity being defined as a maximum value, the intersections of the complexity-boundary time diagrams indicate points where complexity switches from one upper branch to another, known as $\tau_{turning}$ \footnote{The different branches in the CGR arises from $\frac{d\mathcal{C}_{Any}}{d\tau}\sim P_v\sim \sqrt{\tilde{U}}$. The CGR is a multivalued function of $r_{min}$, then there are two branches of the two extremal surfaces (with the same boundary condition and anchoring), but the complexity computed from one of them at some critical time dominates as the same transition is shown in \cite{phase}.}. At this time, a transition occurs in $\mathcal{C}_{Any}$-boundary time diagrams.\\   
\begin{figure}[!h]
	\centering
	\subfloat[]{\includegraphics[width=5cm]{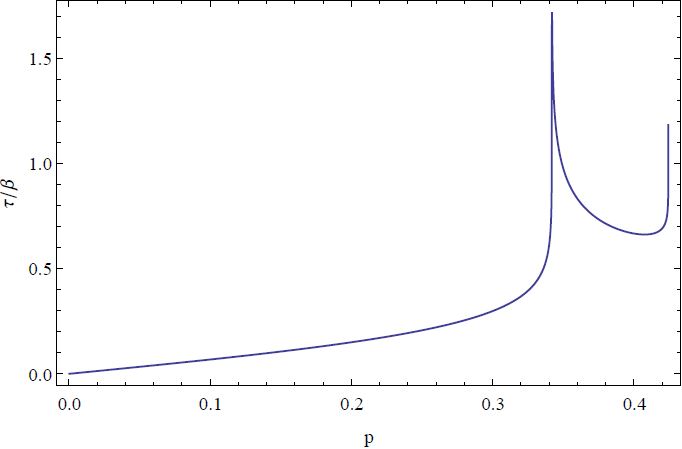} \label{cgr1}}
	\subfloat[]{\includegraphics[width=5cm]{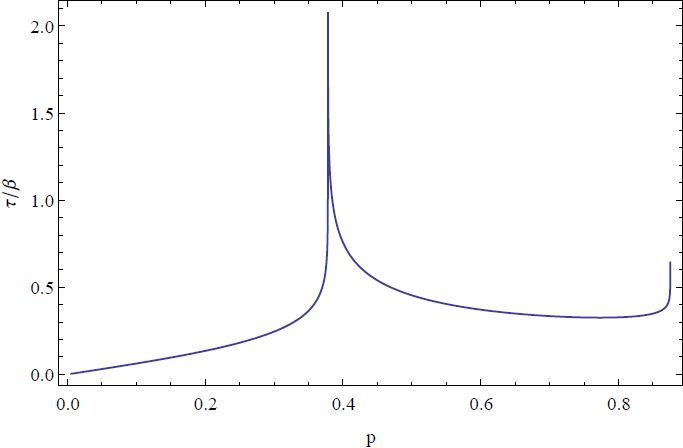} \label{cgr3}}
	\qquad
	\subfloat[]{\includegraphics[width=5cm]{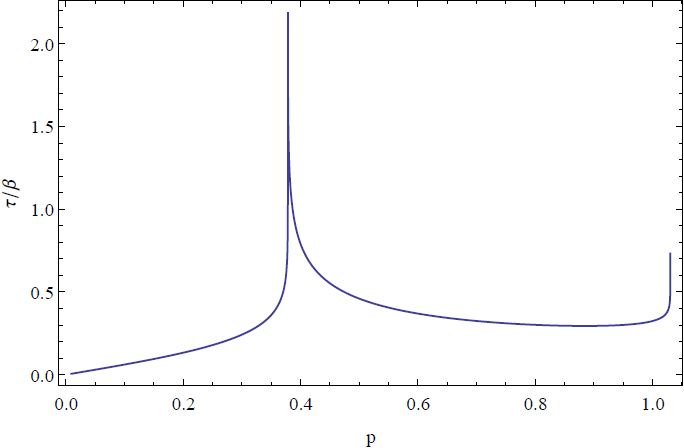} \label{cgr4}}
	\subfloat[]{\includegraphics[width=5cm]{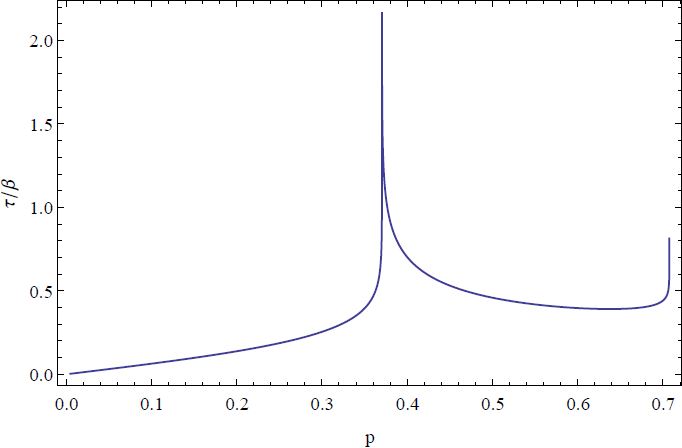} \label{cgr6}}
	\caption{The CGR in terms of the boundary time for $b=C^2$ (a) with the parameters $Q= -0.09$, $q_2=-0.06$ and $\gamma= - 0.002$. (b) $Q=-0.09$, $q_2=-0.03$ and $\gamma=-0.002$. (c) $Q=-0.09$, $q_2=-0.03$ and $\gamma=0.002$. 
(d) $Q=-0.1$, $q_2=-0.03$ and $\gamma=- 0.002$.}\label{fig1}
\end{figure}\\	

\begin{figure}[!h]
	\centering
	\subfloat[]{\includegraphics[width=5cm]{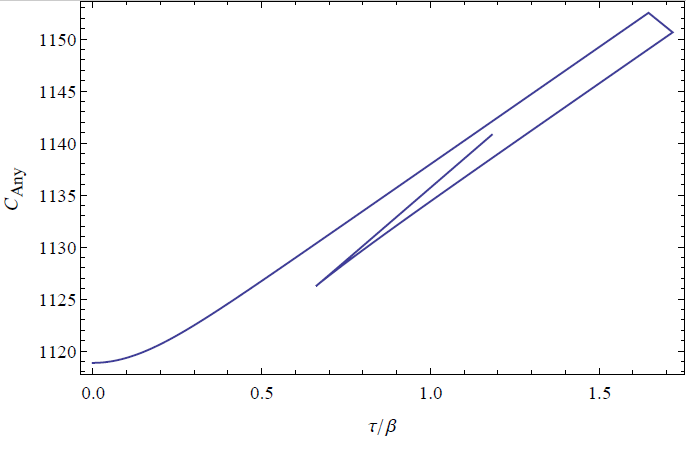} \label{c1}}
	\subfloat[]{\includegraphics[width=5cm]{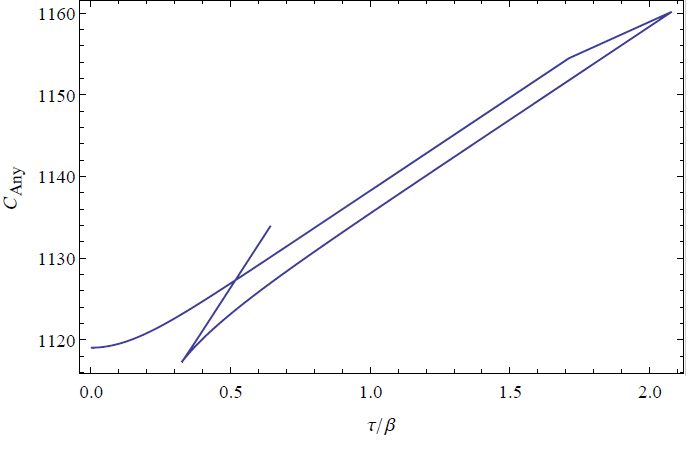} \label{c3}}
	\qquad
	\subfloat[]{\includegraphics[width=5cm]{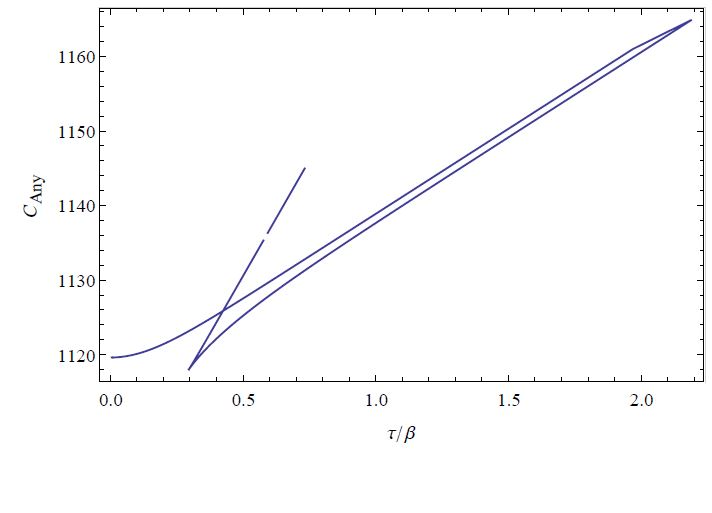} \label{c4}}
	\subfloat[]{\includegraphics[width=5cm]{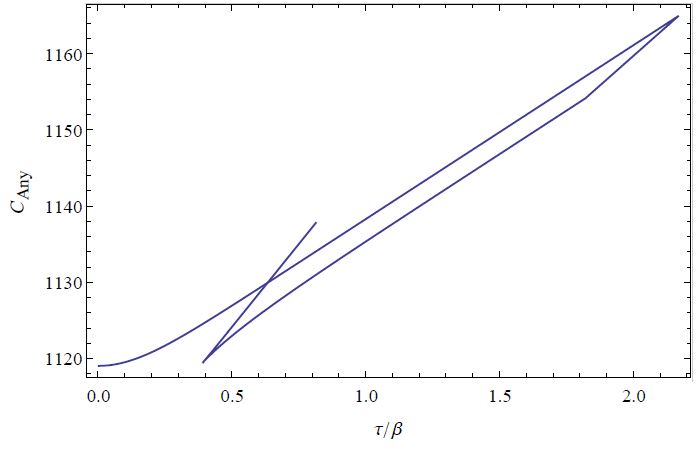} \label{c6}}
	\caption{Complexity in terms of the boundary time for $b=C^2$ (a) with the parameters $Q= -0.09$, $q_2=-0.06$ and $\gamma= - 0.002$. (b) $Q=-0.09$, $q_2=-0.03$ and $\gamma=-0.002$. (c) $Q=-0.09$, $q_2=-0.03$ and $\gamma=0.002$. 
(d) $Q=-0.1$, $q_2=-0.03$ and $\gamma=- 0.002$.}\label{fig2}
\end{figure}

\begin{figure}[!h]
	\centering
	\subfloat[]{\includegraphics[width=5cm]{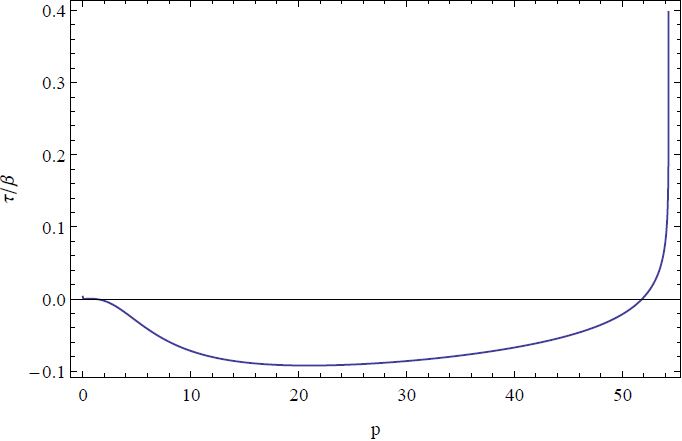} \label{cgrrf1}}
	\subfloat[]{\includegraphics[width=5cm]{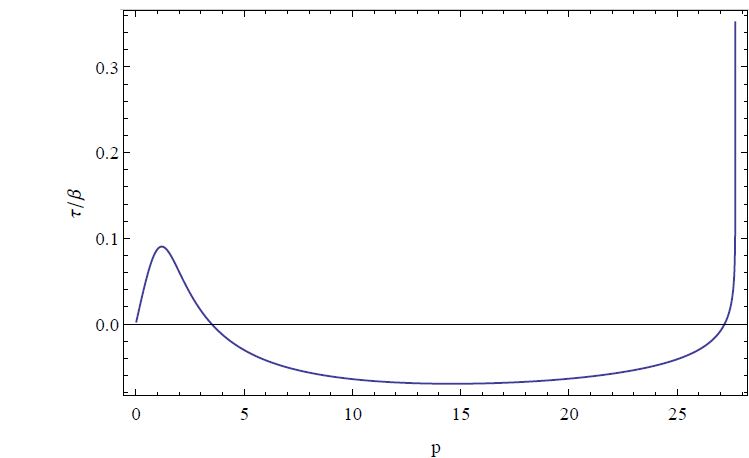} \label{cgrrf3}}
	\qquad
	\subfloat[]{\includegraphics[width=5cm]{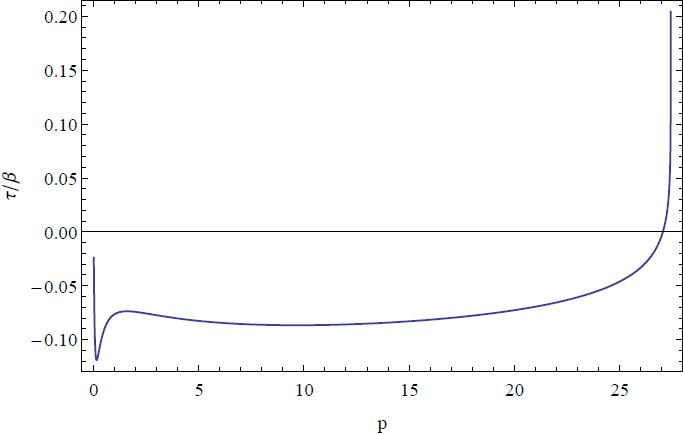} \label{cgrrf4}}
	\subfloat[]{\includegraphics[width=5cm]{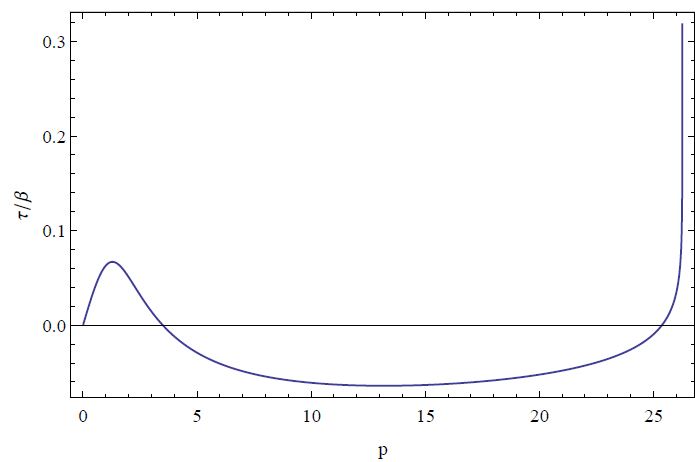} \label{cgrrf6}}
	\caption{The CGR in terms of the boundary time for $b=R^2F^2$ (a) with the parameters $Q= -0.09$, $q_2=-0.06$ and $\gamma= - 0.002$. (b) $Q=-0.09$, $q_2=-0.03$ and $\gamma=-0.002$. (c) $Q=-0.09$, $q_2=-0.03$ and $\gamma=0.002$. 
(d) $Q=-0.1$, $q_2=-0.03$ and $\gamma=- 0.002$.}\label{fig3}
\end{figure}

\begin{figure}[!h]
	\centering
	\subfloat[]{\includegraphics[width=5cm]{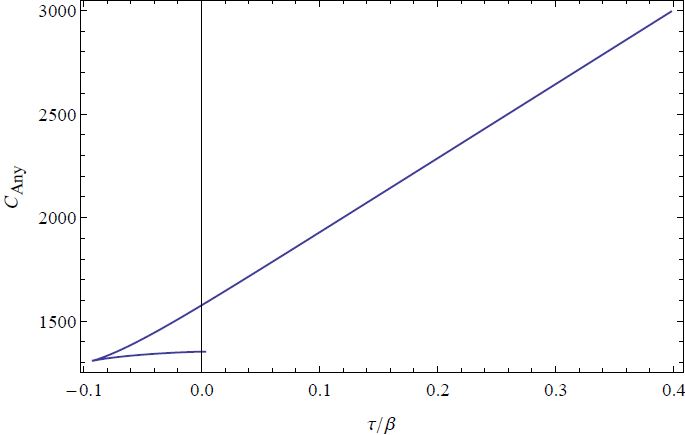} \label{crf1}}
	\subfloat[]{\includegraphics[width=5cm]{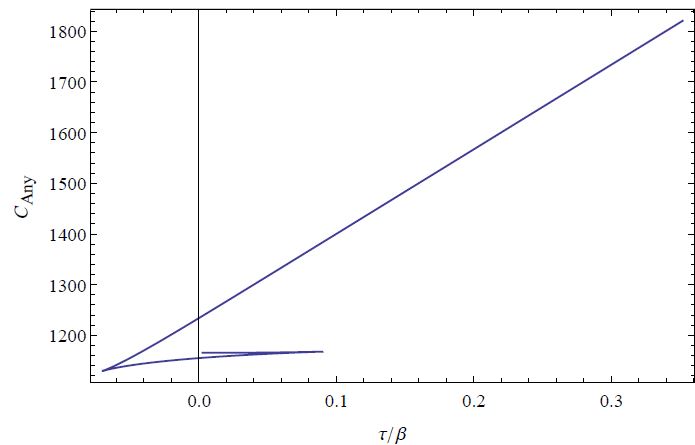} \label{crf3}}
	\qquad
	\subfloat[]{\includegraphics[width=5cm]{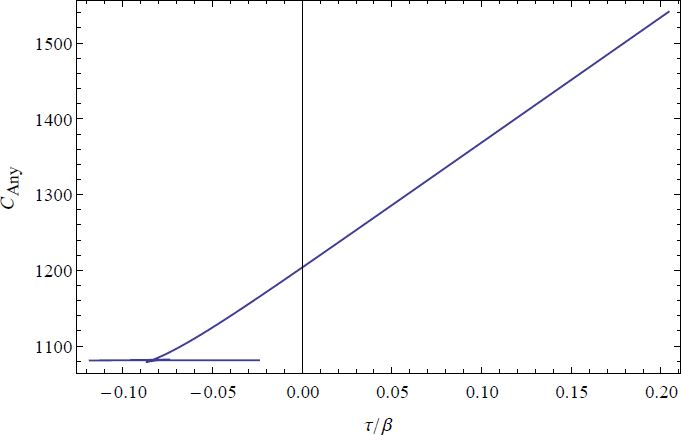} \label{crf4}}
	\subfloat[]{\includegraphics[width=5cm]{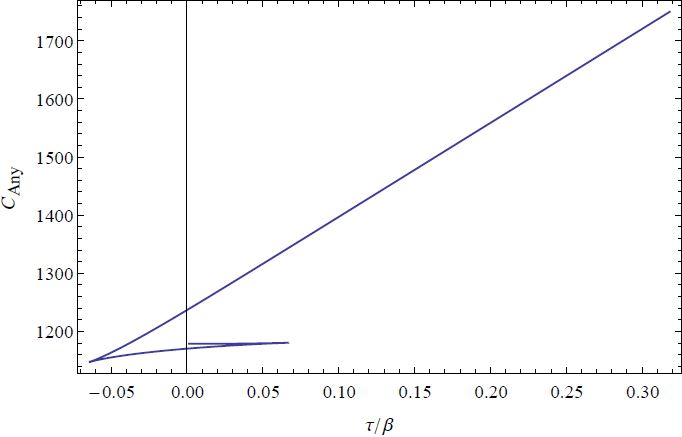} \label{crf6}}
	\caption{Complexity in terms of the boundary time for $b=R^2F^2$ (a) with the parameters $Q= -0.09$, $q_2=-0.06$ and $\gamma= - 0.002$. (b) $Q=-0.09$, $q_2=-0.03$ and $\gamma=-0.002$. (c) $Q=-0.09$, $q_2=-0.03$ and $\gamma=0.002$. 
(d) $Q=-0.1$, $q_2=-0.03$ and $\gamma=- 0.002$.}\label{fig4}
\end{figure}

\begin{figure}[!h]
	\centering
	\subfloat[]{\includegraphics[width=5cm]{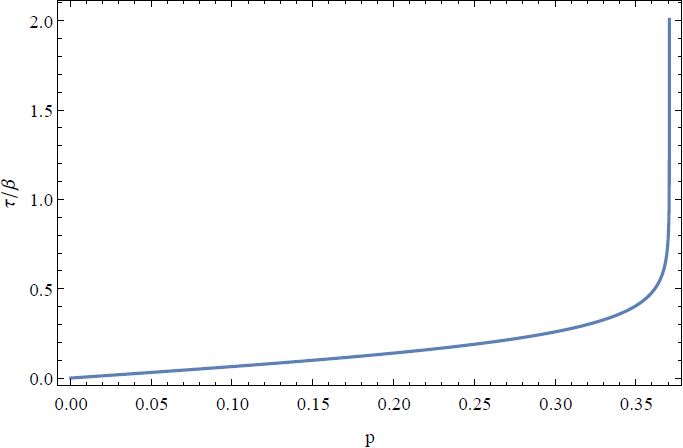} \label{cgrf1}}
	\subfloat[]{\includegraphics[width=5cm]{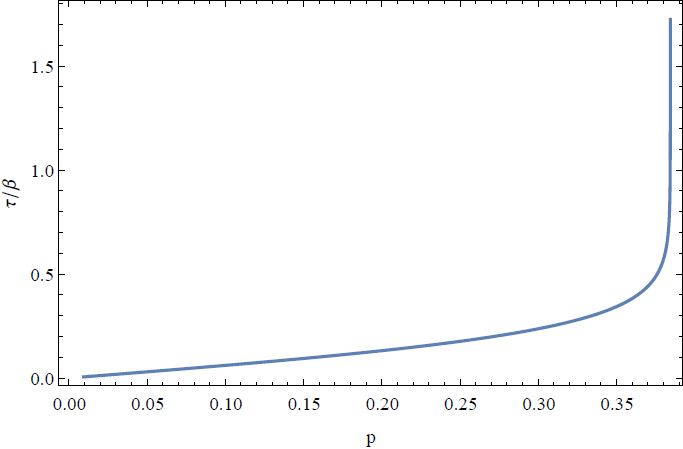} \label{cgrf3}}
	\qquad
	\subfloat[]{\includegraphics[width=5cm]{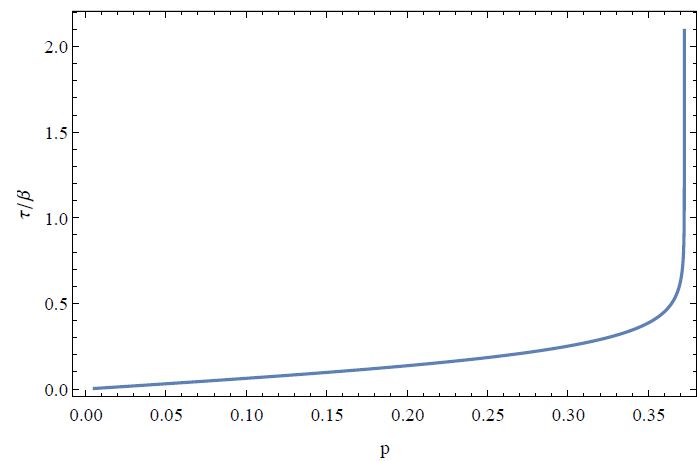} \label{cgrf4}}
	\subfloat[]{\includegraphics[width=5cm]{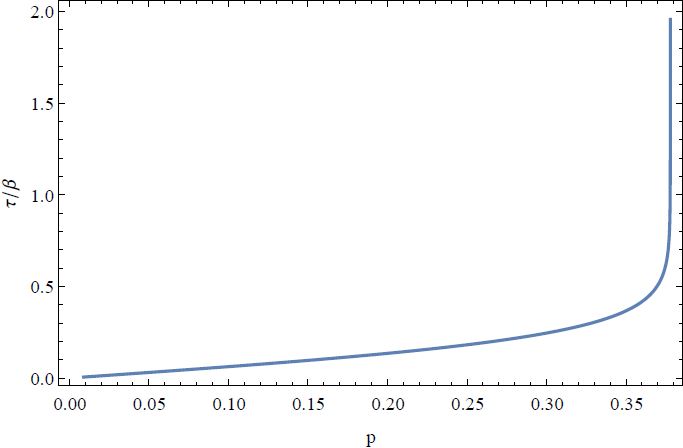} \label{cgrf6}}
	\caption{The CGR in terms of the boundary time for $b=F^2$ (a) with the parameters $Q= -0.09$, $q_2=-0.06$ and $\gamma= - 0.002$. (b) $Q=-0.09$, $q_2=-0.03$ and $\gamma=-0.002$. (c) $Q=-0.09$, $q_2=-0.03$ and $\gamma=0.002$. 
(d) $Q=-0.1$, $q_2=-0.03$ and $\gamma=- 0.002$.}\label{fig5}
\end{figure}

\begin{figure}[!h]
	\centering
	\subfloat[]{\includegraphics[width=5cm]{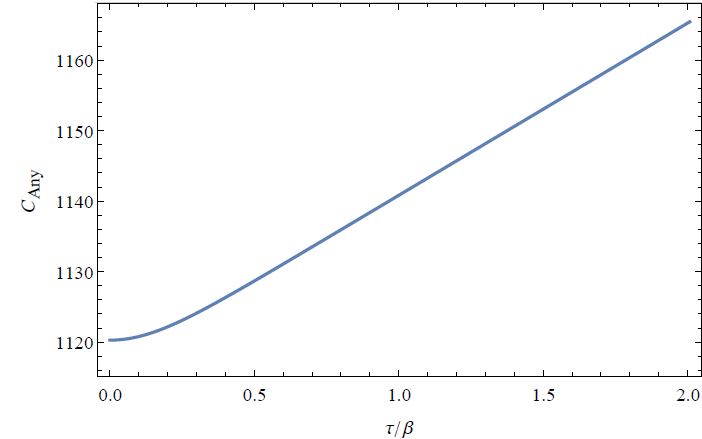} \label{cf1}}
	\subfloat[]{\includegraphics[width=5cm]{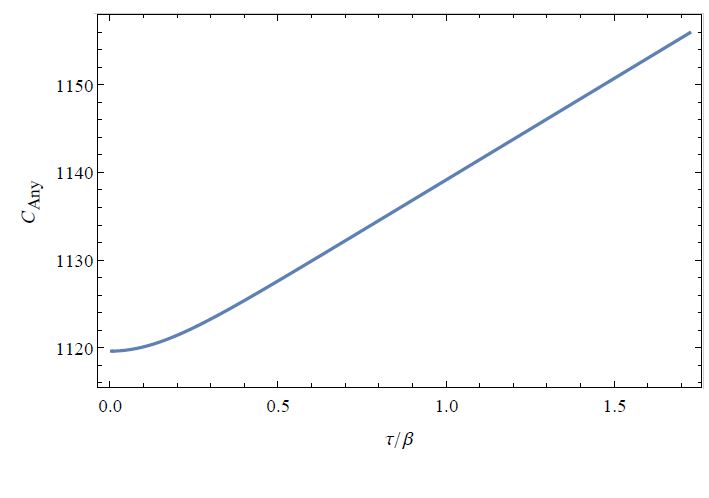} \label{cf3}}
	\qquad
	\subfloat[]{\includegraphics[width=5cm]{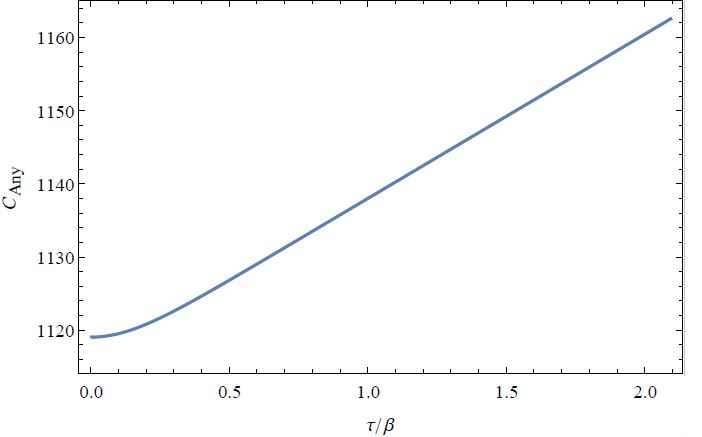} \label{cf4}}
	\subfloat[]{\includegraphics[width=5cm]{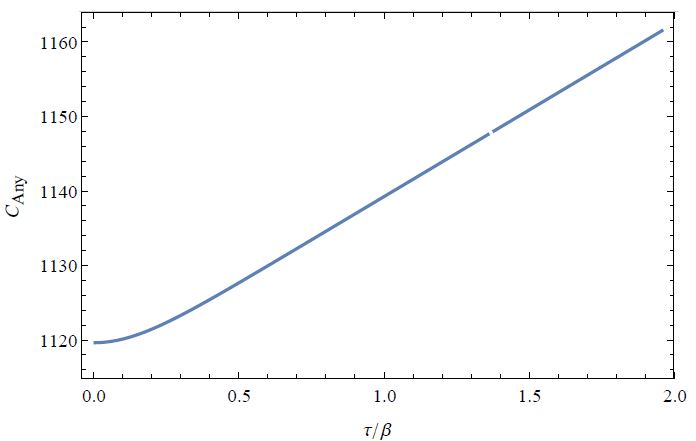} \label{cf6}}
	\caption{Complexity in terms of the boundary time for $b=F^2$ (a) with the parameters $Q= -0.09$, $q_2=-0.06$ and $\gamma= - 0.002$. (b) $Q=-0.09$, $q_2=-0.03$ and $\gamma=-0.002$. (c) $Q=-0.09$, $q_2=-0.03$ and $\gamma=0.002$. 
(d) $Q=-0.1$, $q_2=-0.03$ and $\gamma=- 0.002$.}\label{fig6}
\end{figure}
\section{Physical interpretation of The CGR}\label{sec4}
As shown in Figs.(\ref{fig2}), (\ref{fig4}) and (\ref{fig6}), varying the parameters $Q$, $q_2$ and $\gamma$ leads to noticeable differences in the structure of the complexity growth rate (CGR). In particular, both the locations of the branches and the separation between them depend sensitively on these parameters. A natural question is whether these differences admit a clear physical interpretation.

It was shown in \cite{cse} that, for the choice $b=C^2$, the late time values of the CGR, namely the points at which different branches merge, are controlled by the dynamics of the bulk energy-momentum tensor. For instance, one finds at the location $r_i$ corresponding to the late time of the CGR:
\begin{align}
2\gamma g^{(0)}_{tt}\mathscr{C}^2=r_i^{2}-g^{(4)}_{tt}.\label{divener}
\end{align}
where $\mathscr{C}$ is the Weyl squared tensor evaluated at $r_i$, the location of the late time of the CGR, $g^{(4)}_{tt}\sim <T_{tt}>$, for $T_{tt}$ the component of the energy-momentum tensor and $g_{tt}^{(0)}$ is the $tt$-component of the metric \eqref{metric} in Fefferman-Graham coordinates at the cut-off $r_i$. An analogous reasoning applies to the other choices of the generalization parameter $b$.

In the following subsections, we provide a physical interpretation of the dependence of the CGR on each of the three parameters $\gamma b$, $q_2$, and $Q$.

\subsection{The generalization term $\gamma b$}
We begin by interpreting the generalized term $\gamma b$ as a penalty factor in the dual quantum circuit.

In holographic approaches to quantum complexity, the complexity of a boundary state is computed by extremizing a bulk functional, which defines a geometric structure governing the optimal trajectories associated with state preparation. In Nielsen’s geometric formulation of circuit complexity, this structure is encoded in a cost metric on the space of unitary operators, where penalty factors weight different directions corresponding to distinct classes of quantum gates \cite{cqm}:
\begin{align}
\mathcal{C}_{QFT}=\int dt \sqrt{\sum_{\alpha} \gamma_{\alpha}(Y^{\alpha})^2}=\int ds \sqrt{G_{IJ}\dot{Y}^I\dot{Y}^J}\label{cft}
\end{align}
where the metric is defined in a Hilbert space of unitary operators:
\begin{align}
ds^2=\sum_{\alpha} \gamma_{\alpha}(Y^{\alpha})^2 dt^2.
\end{align}
Here, the coefficients $\gamma_{\alpha}$ act as penalty factors, weighting different directions in the space of gates and suppressing costly operations.

Within the complexity=anything framework, the generalized scalar functional entering the bulk action \eqref{Any} modifies the effective geometry in which the extremal surface or particle trajectory is evaluated:
\begin{align}
\mathcal{C}_{Any}\sim \int d\sigma r^{d-1}a\sqrt{g_{ij}\frac{\partial x^i}{\partial \sigma}\frac{\partial x^j}{\partial \sigma}}=\int d\sigma \sqrt{(r^{2(d-1)}a^2 g_{ij}) \dot{x}^i \dot{x}^j}=\int d\sigma \sqrt{g'_{ij}\dot{x}^i \dot{x}^j} \label{lagran}
\end{align}
where $g'_{ij}=r^{2(d-1)}a^2 g_{ij}$. As shown in \eqref{lagran}, the generalized term enters multiplicatively in the integrand and therefore deforms the metric governing the bulk trajectory. This deformation is fully analogous to the introduction of penalty factors in circuit complexity, where certain directions in the space of unitaries are assigned a higher computational cost.

Importantly, this identification does not rely on specifying a unique set of boundary gates or a microscopic circuit construction, which is known to be intrinsically ambiguous in holographic duality \cite{general}. Rather, the generalized functional encodes a penalty structure at the level of the effective complexity geometry. Different choices of the generalized parameter correspond to inequivalent cost metrics and therefore to different notions of optimal circuit complexity, all of which are consistent within the inherent freedom of defining holographic complexity.

This interpretation is further supported by rewriting the complexity functional in a form directly comparable to Nielsen’s action, where the generalized factor appears as a position-dependent weight in the effective metric, the factor $(r^{d-1}a)^2$ in \eqref{lagran}, which we refer to as a \textit{curvature penalty factor}. In this sense, the generalized term provides a bulk realization of penalty factors that control the relative cost of different bulk trajectories, and consequently modify the resulting CGR.

We emphasize that this penalty-factor interpretation should be understood as a structural correspondence rather than a microscopic identification. The intrinsic ambiguity in the choice of gates and penalties in quantum circuits is mirrored holographically by the freedom to choose the generalized functional, and the analytical relation derived here makes this correspondence explicit.

This analogy can be strengthened using the correspondence between bulk symplectic forms and boundary Berry curvature. In holography, variations of bulk observables are related to the Berry curvature form $\omega$ in the dual theory \cite{general2}, which is itself tied to the imaginary part of the quantum geometric tensor or Fubini–Study (FS) metric \cite{cheng}:
\begin{align}
\delta \mathcal{C}_{Any}=\Omega=\omega=\Im \mathcal{G}_{IJ},
\end{align}
$\Omega$ is the bulk symplectic form and $\omega$ the Berry curvature form. The FS metric can be related to $G_{IJ}$ as:
\begin{align}
G_{IJ}=\mathcal{P}_{IM}\mathcal{G}^{MN}\mathcal{P}_{N J},
\end{align}
where $\mathcal{P}_{IJ}$ encodes the penalty factors. One could rewrite \eqref{cft} in terms of the bulk symplectic form as follows:
\begin{align}
\int ds \sqrt{\mathcal{P}_{IM}\mathcal{G}^{MN}\mathcal{P}_{N J}\dot{Y}^I\dot{Y}^J}&=\int ds \sqrt{\mathcal{P}_{IM}(\Re{\mathcal{G}^{MN}}+i\Im{\mathcal{G}^{MN}})\mathcal{P}_{N J}\dot{Y}^I\dot{Y}^J},\\
&=\int ds \sqrt{\mathcal{P}_{IM}(\Re{\mathcal{G}^{MN}}+i \delta\mathcal{C}_{Any})\mathcal{P}_{NJ}\dot{Y}^I\dot{Y}^J},\\
&=\int ds \sqrt{\mathcal{P}_{IM}(\Re{\mathcal{G}^{MN}}+i \delta\int d\sigma\sqrt{g'_{ij}\dot{x}^i \dot{x}^j})\mathcal{P}_{NJ}\dot{Y}^I\dot{Y}^J}.
\end{align}
Then,
\begin{align}
\mathcal{C}_{QFT}=\int ds \sqrt{\mathcal{P}_{IM}\mathcal{P}_{NJ}\Re{\mathcal{G}^{MN}}\dot{Y}^I\dot{Y}^J+i \mathcal{P}_{IM}\mathcal{P}_{NJ}\delta\int d\sigma\sqrt{g'_{ij}\dot{x}^i \dot{x}^j}\dot{Y}^I\dot{Y}^J}.\label{qfthol}
\end{align}
Expressing the circuit complexity in terms of the bulk functional then shows explicitly that variations of $a(r)$ source deformations of the penalty matrix $\mathcal{P}_{IJ}$. Consequently, modifying $\gamma b$ directly corresponds to deforming the penalty structure of the dual quantum circuit. Given that holography does not uniquely specify a gate set or penalty scheme, any gravitational complexity proposal necessarily corresponds to an equivalence class of quantum circuits. Within this framework, we show analytically that the generalized functional deforms the effective cost metric in a manner identical to the introduction of a penalty factor in Nielsen’s geometric approach. In this precise sense, the generalized term encodes a bulk realization of circuit penalties.

A natural question arises: is there a penalty factor in quantum circuits whose variation alters the corresponding quantities, namely $\mathcal{C}_{Any}$ as the gate complexity and $\dot{\mathcal{C}}_{Any}$ as the circuit depth \cite{3lec}? In the section \ref{qubitcir}, we introduce two such circuit constructions that are applicable to quantum computation.

\subsection{The non-minimal coupling $q_2$}
We now argue that the non-minimal coupling $q_2$ controls an effective scrambling time in the dual theory.

The non-minimal term $R^2F^2$ deforms Einstein-Maxwell theory and hence its holographic dual. Upon dimensional reduction, the undeformed theory is related to the Sachdev-Ye-Kitaev (SYK) model \footnote{We should note that the duality of SYK model and black holes (Jackiew-Teitelboim gravity) is well-established for the $q_2=0$, without the non-minimal coupling. With $q_2\neq 0$ the dual theory would be a deformed SYK model; nonetheless, at low-curvature regime it reduces to SYK model. In addition, our results about the curvature penalty factor is intact whether $q_2=0$ or not.}, which is known to saturate the bound on chaos and exhibit extremely fast scrambling. Introducing $q_2$ moves the system away from this maximally chaotic limit.

In the holographic description of quantum complexity, the growth rate of complexity is controlled not only by the bulk dynamics but also by the geometric structure induced by the complexity functional itself. In particular, within Nielsen’s geometric approach to circuit complexity, the scrambling time is understood as the timescale over which operators spread in the space of unitary transformations, a notion that depends on the choice of cost function and penalty factors.

In the present framework, the generalized complexity functional induces an effective geometry governing the optimal bulk trajectories associated with complexity growth. The presence of the generalized functional effectively rescales the metric entering the particle action \eqref{Any} governing complexity \footnote{It is worth mentioning that this is a scaling and not the transformation of the coordinates.}:
\begin{align}\label{metn}
ds^2_{scal.}:=-a(r)^2 g_{tt}dt^2+a(r)^2g_{rr}dr^2+r^2(dx^2+dy^2),
\end{align}
where $g_{xx}$ and $g_{yy}$ are not scaled. As shown in \eqref{metn}, the generalized term rescales the metric entering the particle action that computes the complexity. Importantly, this rescaling does not alter the bulk equations of motion and therefore does not modify the intrinsic chaotic properties of the black brane background. Instead, it changes the notion of distance relevant for complexity, in direct analogy with the introduction of penalty factors in quantum circuit geometry.

As a consequence, the butterfly velocity \footnote{Butterfly velocity is a measure that how fast information, or the effect of a tiny perturbation, spreads through a quantum system. More precisely, for two local quantum operators $W$ and $V$ it is given by out-of-time-order correlators, $-<[W(t,x),V(0)]^2>$, where at large $t$, its growth is proportional to $e^{\lambda_L(t-\frac{x}{v_B})}$ for the parameter $\lambda_L$, the Lyapunov exponent (how fast chaos spreads spatially).} extracted from the effective metric differs from that associated with the physical spacetime metric. The effective butterfly velocity for the metric \eqref{metn} is computed \cite{dimitri}:
\begin{align}
v_{(scal.)B}^2=\frac{(a^2g_{tt})'}{(d-1)g'_{xx}},
\end{align}
where the derivatives are with respect to $r$. We find that it is rescaled relative to the standard butterfly velocity \footnote{$v_B^2$ is the butterfly velocity of the bulk, the metric \eqref{metric}.}:
\begin{align} 
v_{(scal.)B}^2&=\frac{1}{a^2}\Big|_{r=R=1}v_B^2.
\end{align}
This rescaling leads to an effective scrambling time (for details see Appendix \ref{scram}):
\begin{align}\label{tn}
t_{scal.}\approx \frac{x}{v_{(scal.)B}}=\frac{t_*}{|a(R)|}.
\end{align}
It is important to emphasize that the scrambling time discussed in this section is not the microscopic scrambling time defined via out-of-time-order correlators or shock-wave analyses, which depends solely on the bulk equations of motion. Rather, we introduce an effective scrambling time that governs the rate of complexity growth. This timescale depends on the geometry entering the complexity functional and therefore can be sensitive to the generalized parameter defining the cost metric.

This interpretation is further supported by the near-horizon expansion of the CGR, where increasing $|q_2|$ generically leads to an enhancement of the effective scrambling time and a corresponding suppression of the CGR, in agreement with the numerical results presented below.

In the following it is shown that the CGR near-horizon expansion is inversely related to the scrambling time. The leading term in \eqref{tn} would be:
\begin{align}
t_{scal.}\approx \frac{1}{\sqrt{g'_0(R)}}\Big(1-\frac{q_2}{2}\frac{g'_1(R)}{g'_0(R)}-\gamma b_0(R)-\gamma q_2b_1(R)+\frac{\gamma q_2 b_0(R)}{2}\frac{g'_1(R)}{g'_0(R)}\Big),\label{texp}
\end{align}
where $g_{tt}=g_0+q_2g_1$ and $b=b_0+q_2b_1$. In addition, the leading term of the complexity growth rate in \eqref{cdotf} would be:
\begin{align}
\dot{\mathcal{C}}_{Any}(r_f)&\sim a(r_f)\sqrt{g_{tt}(r_f)},\\
&\approx \sqrt{g_0(r_f)}\Big(1+\frac{q_2}{2}\frac{g_1(r_f)}{g_0(r_f)}+\gamma b_0(r_f)+\gamma q_2b_1(r_f)+\frac{\gamma q_2 b_0(r_f)}{2}\frac{g_1(r_f)}{g'_0(r_f)}\Big).\label{capp}
\end{align}
By the following approximations near the horizon $r_f\approx R$:
\begin{align}
&g'_{tt}(R)(r_f-R)\approx g_{tt}(r_f),\\
&g'_0(R)(r_f-R)\approx g_0(r_f),\\
&g'_1(R)(r_f-R)\approx g_1(r_f).
\end{align}
\eqref{capp} is given:
\begin{align}
\frac{\sqrt{r_f-R}}{\dot{\mathcal{C}}_{Any}(r_f)}\approx \frac{1-\frac{q_2}{2}\frac{g'_1(R)}{g'_0(R)}-\gamma b_0(r_f)-\gamma q_2b_1(r_f)+\frac{\gamma q_2 b_0(r_f)}{2}\frac{g'_1(R)}{g'_0(R)}}{\sqrt{g'_0(R)}}.\label{incdot}
\end{align}
Putting \eqref{incdot} into \eqref{texp}, we could write the new scrambling time in terms of the complexity growth rate:
\begin{align}\label{cdotscr}
t_{scal.}&\approx \frac{\sqrt{r_f-R}}{\dot{\mathcal{C}}_{Any}(r_f)}+\frac{\gamma}{\sqrt{g'_0(R)}}\Big(b'_0(R)+q_2b'_1(R)\Big)(r_f-R)\nonumber \\
&+\frac{\gamma q_2}{\sqrt{g'_0(R)}}\frac{g'_1(R)}{g'_0(R)}\Big(b_0(R)\Big),
\end{align}
where $b'_0(R)(r_f-R)\approx b_0(r_f)-b_0(R)$. Then, the complexity growth rate near the horizon could be rewritten in terms of the bulk scrambling time:
\begin{align}
\dot{\mathcal{C}}_{Any}(r_f)&\approx \frac{\sqrt{r_f-R}}{\frac{t_*}{|a(R)|}-\frac{\gamma}{\sqrt{g'_0(R)}}\Big(b'(R)\Big)(r_f-R)-\frac{\gamma q_2}{\sqrt{g'_0(R)}}\frac{g'_1(R)}{g'_0(R)}\Big(b_0(R)\Big)},\nonumber\\
&\approx \frac{|1+\gamma b(R)|\sqrt{r_f-R}}{t_*-\frac{\gamma}{\sqrt{g'_0(R)}}\Big(b'(R)\Big)(r_f-R)-\frac{\gamma q_2}{\sqrt{g'_0(R)}}\frac{g'_1(R)}{g'_0(R)}\Big(b_0(R)\Big)},\label{cdotscram}\\
&\approx \frac{1}{t_{scal.}}=\frac{|a(R)|}{t_*}.
\end{align}\\
Now, we get into the numerical results.\\

\textbf{The case $b=C^2$}\\
In this case we have:
\begin{align}
&b_0(R)=12-24Q^2+12Q^4,\label{b0}\\
&b'_0(R)=-72+192Q^2-120Q^4,\label{b'0}\\
&g'_0(R)=3-Q^2,\\
&g'_1(R)=1056Q^2.
\end{align}
By \eqref{cdotscram} and setting $\gamma=-0.002$, we have:
\begin{align}
\dot{\mathcal{C}}_{Any}&\approx \frac{1+\gamma q_2}{(1+0.004q_2)(-0.0832509-4.6865 q_2+0.0396679q_2)},\nonumber \\
&\approx 1+q_2\approx 1-|q_2|.
\end{align}
Then, by the increase in $|q_2|$, as a result, there would be an increase in the new scrambling time, the complexity growth rate is decreased as Fig.(\ref{cgr1}) and (\ref{cgr3}) demonstrate.\\

\textbf{The case $b=R^2F^2$}\\
$b_0(R)$, $b'_0(R)$ and $b_1(R)$ for $b=R^2F^2$ would be:
\begin{align}
&b_0(R)=-288Q^2,\label{b0rf}\\
&b'_0(R)=1952Q^2,\\
&b_1(R)=-6144 (54 Q^2 - 84 Q^4 + 21 Q^6) q_2.\label{b1rf}
\end{align}
The complexity growth rate for $b=R^2F^2$ according to \eqref{cdotscram}, \eqref{b0rf}, \eqref{b1rf} and the fact that $\gamma=-0.002$ is given:
\begin{align}
\dot{\mathcal{C}}_{Any}&\approx \frac{1+\gamma (-288Q^2-2653.59 q_2)}{1+0.004q_2-\Big(0.3312 - 12.0417 q2\Big)},\\
&\approx 1-q_2=1+|q_2|
\end{align}
As Fig.s(\ref{cgrrf1}) and (\ref{cgrrf3}) for $Q=-0.09$ show the complexity growth rate is decreased by the increase of $|q_2|$.\\

\textbf{The case $b=F^2$}\\
The leading term in $b_0(R)$, $b'_0(R)$ and $b_1(R)$ for $b=F^2$ would be:
\begin{align}
&b_0(R)=-2Q^2,\label{b0f}\\
&b'_0(R)=8Q^2,\label{b'0f}\\
&b_1(R)=1280 Q^4 - 2304 Q^2.\label{b1f}
\end{align}
The leading order in the CGR based on \eqref{cdotscram}, \eqref{b0f},  \eqref{b'0f} and setting $\gamma=-0.002$ would be:
\begin{align}
\dot{\mathcal{C}}_{Any}&\approx \frac{1+\gamma (-2Q^2-2304 q_2)}{1+0.004q_2-\gamma \Big(0.0374629 - 10.7893 q2\Big)},\\
&\approx 1+q_2=1-|q_2|
\end{align}
Fig.s (\ref{cgrf3}) and (\ref{cgrf6}) illustrate that the CGR is reduced by the increase in the non-minimal coupling.

In this section we showed that the new scrambling time is related to the non-minimal coupling $q_2$ and consequently, by the change of the coupling, the CGR is changed as is expected.

\subsection{The conserved charge $Q$}
Finally, we interpret the conserved charge $Q$ and argue that it effectively restricts the set of simple unitary operators available in the dual quantum circuit. As a result, increasing $Q$ suppresses the complexity growth rate.

A useful way to understand this effect comes from the Solovay-Kitaev algorithm (SKA), which guarantees that an arbitrary unitary operator $U$ in a finite-dimensional Hilbert space can be approximated by a sequence of simple gates drawn from a fixed universal set, i.e. $U\approx g_1 g_2...$ where $g_i$s are some simple unitary operators. The accuracy of this approximation depends on the distance between $U$ and the identity operator. Writing $U=e^{iH}$, this distance is bounded as $d=max [2\sin (|E|/2)]\le ||H||$, where $||H||$ is the norm of the Hermitian operator and $E$, eigenvalues \cite{nielsen}.

In short, SKA states that for an arbitrary accuracy $\epsilon$, if $d(I,U)<\epsilon$, then there exist $V$ and $W$ such that $d(U,VWV^{\dagger} W^{\dagger})<\epsilon^{3/2}c_1$ and $d(I,V),d(I,W)<\sqrt{\epsilon}c_2$ where $I$ is identity operator, $c_1$ and $c_2$ are constant \cite{nielsen}. In addition, it can be shown \cite{kitaev} that if $V=e^{iF}$ and $W=e^{iG}$, then:
\begin{enumerate}
\item $[F,G]=iH$,\\
\item $||F||,||G||\le \sqrt{||H||}$. 
\end{enumerate}
These relations explicitly demonstrate that the size of $||H||$ restricts the set of admissible generators for the simple unitary operators.

From the holographic perspective, the conserved charge $Q$ plays an analogous role. Increasing $Q$ effectively increases the scale associated with the generator $H$, thereby reducing the set of allowed simple unitaries in the boundary theory. This restriction manifests itself as a suppression of the rate at which complexity can grow.

This qualitative picture is supported by the bulk analysis. As discussed in the previous subsection, the effective scrambling time increases with $Q$. Since complexity growth is inversely related to the scrambling time, a larger charge leads to a smaller complexity growth rate. Explicitly, in the following subsections near the horizon one finds for all three choices of the generalization parameter that:
\begin{align}
\dot{\mathcal{C}}_{Any}\sim 1-\frac{Q^2}{2}+...,
\end{align}
indicating a decrease of the CGR as $Q$ increases. This behavior is in full agreement with the numerical results shown in Figs.(\ref{fig1}), (\ref{fig3}), and (\ref{fig5}).\\

\textbf{The case $b=C^2$}\\
The complexity growth rate based on \eqref{cdotscram}, \eqref{b0} and \eqref{b'0} is given by:
\begin{align}
\dot{\mathcal{C}}_{Any}(r_f)&\approx \frac{|1+\gamma 12(1-2Q^2+Q^4)|\sqrt{r_f-R}}{t_*-\frac{2\gamma}{\sqrt{3}}(-72+180Q^2)(r_f-R)-\frac{\gamma q_2 (1056Q^2)}{(3-Q^2)^{\frac{3}{2}}}(12-24Q^2)},\\
&\approx \frac{1-2\gamma Q^2}{1+\frac{Q^2}{2}}\approx 1-\frac{Q^2}{2}.
\end{align}
Then, by increase in the conserved charge we see a decrease in the complexity growth rate.\\

\textbf{The case $b=R^2F^2$}\\
The complexity growth rate \eqref{cdotscram} would be:
\begin{align}
\dot{\mathcal{C}}_{Any}(r_f)&\approx \frac{1-288\gamma Q^2}{1+\frac{Q^2}{2}}\approx 1-\frac{Q^2}{2}
\end{align}
Fig.s (\ref{cgrrf3}) and (\ref{cgrrf6}) displays that by increase in the conserved charge, the CGR is reduced.\\

\textbf{The case $b=F^2$}\\
The CGR according to \eqref{cdotscram} is given by:
\begin{align}
\dot{\mathcal{C}}_{Any}(r_f)&\approx \frac{1-2304\gamma Q^2}{1+\frac{Q^2}{2}}\approx 1-\frac{Q^2}{2}
\end{align}
Fig.s (\ref{cgrf3}) and (\ref{cgrf6}) depict the decrease of the CGR by the increase in the conserved charge.

In summary, the conserved charge $Q$ restricts the available simple unitary operators in the dual circuit, thereby increasing the effective scrambling time and suppressing the complexity growth rate, in accordance with both the Solovay-Kitaev framework and the holographic analysis.
\subsection{Superconducting qubits circuits}\label{qubitcir}
In this section, we provide two examples in superconducting qubits circuits. These examples demonstrate that modifying a penalty parameter can either suppress or enhance the effective processing rate, depending on how it constrains the circuit dynamics. This behavior closely parallels the dependence of the holographic complexity growth rate (CGR) on the generalization parameter $\gamma b$.\\

\textbf{Example I: Crosstalk penalty and increased circuit depth.}
In the superconducting qubit circuits to prepare a qubit one needs controlling quickly the fast gates (or low runtime) to stay away from decoherence. However, fast controlling pose a problem because of the fact that it requires a strong, sharp microwave pulse, but strong pulse has a wide frequency spectrum. A strong pulse allows for very fast $CNOT$ gates \cite{guide}. Nonetheless, this makes parallelizing operations to reduce circuit depth. Moreover, a wide frequency pulse excite the qubit to higher energy states which leads to the leakage error or crosstalk between energy levels. To reduce the leakage it requires the qubit be more anharmonic that means the energy gap of the states be more significantly different \cite{guide}. The anharmonicity makes the states more isolated from the states with higher energies. A problem back of the other, the anharmonicity makes the qubit more sensitive to charge noise which results in reducing the coherence time. As a way out, one needs increase the circuit depth by applying crosstalk penalty.

Increase in crosstalk penalty, $\gamma_{\alpha}$ means that the algorithm weights heavily the parallel operations which leads to the crosstalk and makes it sequentially instead, as a consequence, increase in the circuit depth, $\dot{\mathcal{C}}_{circ.}$. To avoid the possible errors, insertion of correction pulses lets to mitigate the effects of errors which increases the gate count, $\mathcal{C}_{circ.}$. In the following we address a concrete example of the crosstalk penalty.

\begin{figure}[!h]
	\centering
	\subfloat[]{\includegraphics[width=8cm]{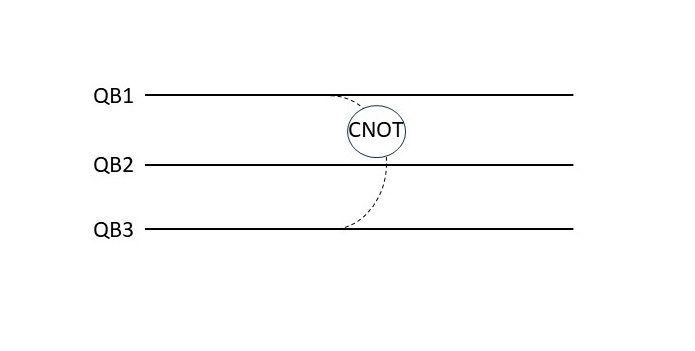} \label{cnot}}
	\subfloat[]{\includegraphics[width=8cm]{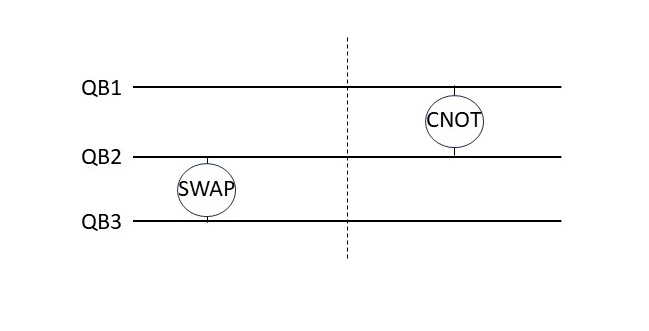} \label{swap}}
	\caption{A schematic quantum circuit with three qubits. The time goes from left to right and each horizontal line depicts a qubit. In addition, the dotted line containing a circle shows an operation or quantum gate and the simple vertical dotted line, the time step. (a) a circuit that applies a CNOT operation to qubit 1 and 3 in one step of time. (b) a circuit that operates a SWAP gate to qubit 2 and 3 in a time step and then applies a CNOT gate to qubit 1 and 3 in the next time step.}\label{fig7}
\end{figure}

Suppose that there are three qubits of QB1 connected to QB2 and QB2 connected to QB3 where crosstalk is high between QB1$\&$QB2 and QB2$\&$QB3. If one needs to apply a CNOT(1,3) operator, it is not possible due to the lack of direct coupling. Instead, one must first apply a SWAP(2,3) operation, followed by CNOT(1,3). Since, SWAP(2,3) gate decomposes into three CNOT gates, this procedure significantly increases the circuit depth, as illustrated in figure (\ref{fig7}). Similar constraints arise in current IBM superconducting quantum processors \cite{ibm}.

In this setting, increasing the crosstalk penalty directly suppresses the processing rate. This behavior mirrors the holographic result in which increasing the penalty factor $\gamma b$ reduces the complexity growth rate.
\\

\textbf{Example II: Penalty terms in quantum approximate optimization algorithms.} A contrasting example is provided by quantum approximate optimization algorithms (QAOA), where penalty terms can be used to enhance computational efficiency rather than suppress it. In these algorithms, constraints are incorporated into the problem Hamiltonian as:
\begin{align}
H_{optimized}=H_{system}+\gamma_a H_{constraints},
\end{align}
where $\gamma_a$ controls the strength of the constraint (similar to $\gamma$ in the generalized complexity). Increasing $\gamma_a$ energetically penalizes states that violate the constraints, effectively guiding the system toward the desired subspace of solutions. As a result, the target state can be reached using fewer layers of quantum gates. In practice, this leads to a reduction in circuit depth, which is particularly advantageous in noisy superconducting devices where long circuits suffer from decoherence. Explicit examples of this behavior have been demonstrated in \cite{compare} where intensifying the obeying rules, the system in shorter steps gets to the desired solution.

Related optimization strategies have also been proposed in \cite{mixer},  known as mixer-phaser ansatz in superconducting processors in which they did not used a penalty-based algorithm. In a layman's term, if the problem system goes away from your desired solution, the penalty factor penalizes the system to obey the constraints. However, in \cite{mixer} the constraints have been inserted before the system circuit. Then, they get rid of the penalizing.

\subsection{Relation to holographic complexity}
These two examples illustrate that penalty factors in quantum circuits can have qualitatively different effects on the processing rate, depending on how they constrain the dynamics. In the holographic framework, this behavior is reflected in the dependence of the CGR on the choice of generalization parameter $b$. Certain choices lead to a suppression of the CGR, while others enhance it, in direct analogy with the superconducting-circuit examples discussed above.

Finally, we note that these circuit-based interpretations can be extended beyond qubit systems. SYK model, which describes strongly interacting fermions and serves as a holographic dual of near-$AdS_2$ gravity \cite{sach}, can be mapped to qubit systems using the Bravyi-Kitaev \cite{fqm} and Jordan-Wigner transformations \cite{jwt}. This correspondence allows the circuit-based intuition developed here to be applied equally to strange metal systems.
\section{Conclusion}\label{conclusion}
In this work, we studied holographic complexity in four-dimensional Einstein-Maxwell theory with a non-minimal coupling of the form $R^2F_{\mu \nu }F^{\mu \nu}$. Since an exact black brane solution is not available, we constructed a perturbative AdS solution to first order in the non-minimal coupling parameter $q_2$. This model exhibits a linear temperature dependence of the resistivity and therefore provides a holographic realization of strange metal behavior, although no phase transition is present.

We analyzed the complexity of this system within the $complexity=anything$ framework, which allows for a generalized definition of holographic complexity. In this formulation, the complexity growth rate (CGR) depends on three independent parameters: the conserved charge $Q$, the non-minimal coupling $q_2$, and the choice of the generalized term $b$ entering the complexity functional. We explicitly considered three representative choices, namely the Weyl tensor squared $C^2$, $R^2F^2$, and the Maxwell invariant $F^2$.

We showed analytically that the generalized bulk functional deforms the effective geometry governing the extremal trajectories associated with complexity. This deformation modifies the effective cost metric in a manner structurally equivalent to the introduction of penalty factors in Nielsen’s geometric approach to quantum circuit complexity. Given the intrinsic ambiguity in the choice of gate sets and penalty schemes in holographic duality, this correspondence should be understood at the level of complexity geometry rather than as a microscopic identification of boundary gates. In this sense, the generalized parameter encodes a bulk realization of penalty structures that control the relative cost of different directions in complexity space.

We further examined the implications of this deformation for the CGR. By analyzing the effective butterfly velocity associated with the deformed complexity geometry, we identified an effective scrambling time that governs the CGR. We demonstrated analytically that this timescale depends on both the non-minimal coupling and the generalized parameter, while the intrinsic chaotic properties of the background geometry remain unchanged. Our numerical results support this analytical picture and show that increasing the strength of the generalized deformation, $|q_2|$ typically leads to a suppression of the CGR through an enhancement of the effective scrambling time.

Moreover, the conserved charge restricts the set of admissible simple unitary operators in the boundary theory, in accordance with the Solovay-Kitaev framework, and consequently suppresses the CGR.

To further support our interpretations, we discussed analogies with superconducting-qubit circuits, where penalty factors can either suppress or enhance the effective processing rate depending on how they constrain circuit dynamics. These examples demonstrate that the holographic behavior of the CGR is consistent with well-established mechanisms in realistic quantum computational platforms.

We found that the behavior of the CGR under variations of $q_2$ depends sensitively on the choice of the generalization parameter. For $b=C^2$ and $b=F^2$, increasing $q_2$ leads to a suppression of the CGR, whereas for $b=R^2F^2$ the opposite behavior is observed. Moreover, the generalization $b=F^2$ does not give rise to multiple branches in the CGR and closely resembles the behavior of the standard CV and CA proposals.

Several directions for future work are suggested by our results. It would be interesting to formulate a holographic complexity proposal that avoids explicit penalty factors and instead constrains the target state directly. Additionally, the appearance of multiple branches in the CGR for curvature-dependent generalizations, but not for spacetime-independent choices such as $F^2$, merits further investigation. Understanding this distinction may shed light on the geometric origin of complexity in holography.
\section*{Acknowledgment}
Authors would like to express their gratitude to Monireh Emami for her valuable help in the numerical calculations. Additionally, OpenAI’s ChatGPT was used to enhance the clarity of this manuscript.
\appendix
\renewcommand\theequation{\thesection-\arabic{equation}} 
\setcounter{equation}{0}
\section{Equations of motion}\label{app1}

The $tt$-component of the equations of motion is as follows:
\begin{eqnarray}\label{tt-comp}
 &&20 q_2 f(r)^3 h'(r)^2 H'(r)^4+28 q_2 f(r)^2 f'(r) h'(r)^2 H'(r)^3+\frac{16 q_2 f(r)^3 h'(r)^2 H'(r)^3}{r}\nonumber\\&&+64 q_2 f(r)^3
h'(r) h''(r) H'(r)^3-35 q_2 f(r) f'(r)^2 h'(r)^2 H'(r)^2\nonumber\\&&-\frac{160 q_2 f(r)^2 f'(r) h'(r)^2 H'(r)^2}{r}-\frac{32 q_2 f(r)^3
	h'(r)^2 H'(r)^2}{r^2}\nonumber\\&&+16 q_2 f(r)^3 h''(r)^2 H'(r)^2-36 q_2 f(r)^2 h'(r)^2 f''(r) H'(r)^2\nonumber\\&&-56 q_2 f(r)^2 f'(r) h'(r) h''(r)
H'(r)^2-\frac{96 q_2 f(r)^3 h'(r) h''(r) H'(r)^2}{r}\nonumber\\&&+72 q_2 f(r)^3 h'(r)^2 H''(r) H'(r)^2+16 q_2 f(r)^3 h'(r) h^{(3)}(r)
H'(r)^2\nonumber\\&&+\frac{68 q_2 f(r) f'(r)^2 h'(r)^2 H'(r)}{r}+\frac{152 q_2 f(r)^2 f'(r) h'(r)^2 H'(r)}{r^2}-24 q_2 f(r)^2 f'(r)
h''(r)^2 H'(r)\nonumber\\&&-\frac{32 q_2 f(r)^3 h''(r)^2 H'(r)}{r}+16 q_2 f(r) f'(r) h'(r)^2 f''(r) H'(r)+\frac{64 q_2 f(r)^2 h'(r)^2
	f''(r) H'(r)}{r}\nonumber\\&&-12 q_2 f(r) f'(r)^2 h'(r) h''(r) H'(r)+\frac{64 q_2 f(r)^3 h'(r) h''(r) H'(r)}{r^2}-16 q_2 f(r)^2 h'(r)
f''(r) h''(r) H'(r)\nonumber\\&&-108 q_2 f(r)^2 f'(r) h'(r)^2 H''(r) H'(r)-\frac{144 q_2 f(r)^3 h'(r)^2 H''(r) H'(r)}{r}+4 q_2 f(r)^2
h'(r)^2 f^{(3)}(r) H'(r)\nonumber\\&&-24 q_2 f(r)^2 f'(r) h'(r) h^{(3)}(r) H'(r)-\frac{32 q_2 f(r)^3 h'(r) h^{(3)}(r) H'(r)}{r}-16 q_2
f(r)^3 h'(r)^2 H^{(3)}(r) H'(r)\nonumber\\&&-\frac{36 q_2 f(r) f'(r)^2 h'(r)^2}{r^2}+\frac{1}{4} q_1 f(r) h'(r)^2-\frac{48 q_2 f(r)^2 f'(r)
	h'(r)^2}{r^3}+\frac{12 q_2 f(r)^3 h'(r)^2}{r^4}\nonumber\\&&-3 q_2 f(r) h'(r)^2 f''(r)^2+\frac{32 q_2 f(r)^2 f'(r) h''(r)^2}{r}+8 q_2
f(r)^2 f''(r) h''(r)^2+\frac{16 q_2 f(r)^3 h''(r)^2}{r^2}\nonumber\\&&-12 q_2 f(r)^3 h'(r)^2 H''(r)^2+\frac{e^{-2 H(r)} \Lambda 
	f(r)}{\mathit{k}}+\frac{e^{-2 H(r)} f(r) f'(r)}{\mathit{k} r}-\frac{12 q_2 f(r) f'(r) h'(r)^2 f''(r)}{r}\nonumber\\&&+\frac{16 q_2 f(r) f'(r)^2
	h'(r) h''(r)}{r}+\frac{40 q_2 f(r)^2 f'(r) h'(r) h''(r)}{r^2}-\frac{32 q_2 f(r)^3 h'(r) h''(r)}{r^3}\nonumber\\&&+4 q_2 f(r) f'(r) h'(r)
f''(r) h''(r)+\frac{80 q_2 f(r)^2 h'(r) f''(r) h''(r)}{r}-10 q_2 f(r) f'(r)^2 h'(r)^2 H''(r)\nonumber\\&&-\frac{8 q_2 f(r)^2 f'(r) h'(r)^2
	H''(r)}{r}+\frac{32 q_2 f(r)^3 h'(r)^2 H''(r)}{r^2}-16 q_2 f(r)^3 h''(r)^2 H''(r)\nonumber\\&&-12 q_2 f(r)^2 h'(r)^2 f''(r) H''(r)-88
q_2 f(r)^2 f'(r) h'(r) h''(r) H''(r)-\frac{96 q_2 f(r)^3 h'(r) h''(r) H''(r)}{r}\nonumber\\&&+2 q_2 f(r) f'(r) h'(r)^2 f^{(3)}(r)+\frac{24
	q_2 f(r)^2 h'(r)^2 f^{(3)}(r)}{r}+16 q_2 f(r)^2 h'(r) h''(r) f^{(3)}(r)\nonumber\\&&+\frac{32 q_2 f(r)^2 f'(r) h'(r) h^{(3)}(r)}{r}+\frac{16
	q_2 f(r)^3 h'(r) h^{(3)}(r)}{r^2}+8 q_2 f(r)^2 h'(r) f''(r) h^{(3)}(r)\nonumber\\&&-16 q_2 f(r)^3 h'(r) H''(r) h^{(3)}(r)-32 q_2
f(r)^2 f'(r) h'(r)^2 H^{(3)}(r)-\frac{32 q_2 f(r)^3 h'(r)^2 H^{(3)}(r)}{r}\nonumber\\&&-32 q_2 f(r)^3 h'(r) h''(r) H^{(3)}(r)+4 q_2 f(r)^2
h'(r)^2 f^{(4)}(r)-8 q_2 f(r)^3 h'(r)^2 H^{(4)}(r)\nonumber\\&&+\frac{e^{-2 H(r)} f(r)^2}{\mathit{k} r^2}=0,
\end{eqnarray}
The $rr$-component of the field equations of motion is as follows,
\begin{eqnarray}\label{rr-comp}
&&-112 e^{2 H(r)} q_2 \mathit{k} f(r)^2 h'(r)^2 H'(r)^4 r^4+240 e^{2 H(r)} q_2 \mathit{k} f(r) f'(r) h'(r)^2
H'(r)^3 r^4+e^{2 H(r)} q_1 \mathit{k} h'(r)^2 r^4\nonumber\\&&-140 e^{2 H(r)} q_2 \mathit{k} f'(r)^2 h'(r)^2 H'(r)^2
r^4-12 e^{2 H(r)} q_2 \mathit{k} h'(r)^2 f''(r)^2 r^4-48 e^{2 H(r)} q_2 \mathit{k} f(r)^2 h'(r)^2 H''(r)^2
r^4\nonumber\\&&+4 \Lambda  r^4-32 e^{2 H(r)} q_2 \mathit{k} f(r) h'(r)^2 H'(r)^2 f''(r) r^4+64 e^{2 H(r)} q_2 \mathit{k}
f'(r) h'(r)^2 H'(r) f''(r) r^4\nonumber\\&&-64 e^{2 H(r)} q_2 \mathit{k} f(r)^2 h'(r) H'(r)^3 h''(r) r^4+128 e^{2 H(r)}
q_2 \mathit{k} f(r) f'(r) h'(r) H'(r)^2 h''(r) r^4\nonumber\\&&-48 e^{2 H(r)} q_2 \mathit{k} f'(r)^2 h'(r) H'(r) h''(r)
r^4+16 e^{2 H(r)} q_2 \mathit{k} f'(r) h'(r) f''(r) h''(r) r^4\nonumber\\&&-32 e^{2 H(r)} q_2 \mathit{k} f(r) h'(r)
H'(r) f''(r) h''(r) r^4-40 e^{2 H(r)} q_2 \mathit{k} f'(r)^2 h'(r)^2 H''(r) r^4\nonumber\\&&+96 e^{2 H(r)} q_2 \mathit{k}
f(r)^2 h'(r)^2 H'(r)^2 H''(r) r^4-64 e^{2 H(r)} q_2 \mathit{k} f(r) f'(r) h'(r)^2 H'(r) H''(r) r^4\nonumber\\&&+48 e^{2
	H(r)} q_2 \mathit{k} f(r) h'(r)^2 f''(r) H''(r) r^4-32 e^{2 H(r)} q_2 \mathit{k} f(r) f'(r) h'(r) h''(r)
H''(r) r^4\nonumber\\&&+64 e^{2 H(r)} q_2 \mathit{k} f(r)^2 h'(r) H'(r) h''(r) H''(r) r^4+8 e^{2 H(r)} q_2 \mathit{k}
f'(r) h'(r)^2 f^{(3)}(r) r^4\nonumber\\&&-16 e^{2 H(r)} q_2 \mathit{k} f(r) h'(r)^2 H'(r) f^{(3)}(r) r^4-16 e^{2 H(r)}
q_2 \mathit{k} f(r) f'(r) h'(r)^2 H^{(3)}(r) r^4\nonumber\\&&+32 e^{2 H(r)} q_2 \mathit{k} f(r)^2 h'(r)^2 H'(r)
H^{(3)}(r) r^4+384 e^{2 H(r)} q_2 \mathit{k} f(r)^2 h'(r)^2 H'(r)^3 r^3\nonumber\\&&-608 e^{2 H(r)} q_2 \mathit{k} f(r)
f'(r) h'(r)^2 H'(r)^2 r^3+4 f'(r) r^3\nonumber\\&&+272 e^{2 H(r)} q_2 \mathit{k} f'(r)^2 h'(r)^2 H'(r) r^3-8 f(r) H'(r)
r^3-48 e^{2 H(r)} q_2 \mathit{k} f'(r) h'(r)^2 f''(r) r^3\nonumber\\&&-32 e^{2 H(r)} q_2 \mathit{k} f(r) h'(r)^2 H'(r)
f''(r) r^3+256 e^{2 H(r)} q_2 \mathit{k} f(r)^2 h'(r) H'(r)^2 h''(r) r^3\nonumber\\&&+64 e^{2 H(r)} q_2 \mathit{k}
f'(r)^2 h'(r) h''(r) r^3-384 e^{2 H(r)} q_2 \mathit{k} f(r) f'(r) h'(r) H'(r) h''(r) r^3\nonumber\\&&+64 e^{2 H(r)}
q_2 \mathit{k} f(r) h'(r) f''(r) h''(r) r^3-32 e^{2 H(r)} q_2 \mathit{k} f(r) f'(r) h'(r)^2 H''(r) r^3\nonumber\\&&-64
e^{2 H(r)} q_2 \mathit{k} f(r)^2 h'(r)^2 H'(r) H''(r) r^3-128 e^{2 H(r)} q_2 \mathit{k} f(r)^2 h'(r) h''(r)
H''(r) r^3\nonumber\\&&+32 e^{2 H(r)} q_2 \mathit{k} f(r) h'(r)^2 f^{(3)}(r) r^3-64 e^{2 H(r)} q_2 \mathit{k} f(r)^2
h'(r)^2 H^{(3)}(r) r^3\nonumber\\&&-144 e^{2 H(r)} q_2 \mathit{k} f'(r)^2 h'(r)^2 r^2-512 e^{2 H(r)} q_2 \mathit{k} f(r)^2
h'(r)^2 H'(r)^2 r^2+4 f(r) r^2\nonumber\\&&+512 e^{2 H(r)} q_2 \mathit{k} f(r) f'(r) h'(r)^2 H'(r) r^2+96 e^{2 H(r)}
q_2 \mathit{k} f(r) h'(r)^2 f''(r) r^2\nonumber\\&&+288 e^{2 H(r)} q_2 \mathit{k} f(r) f'(r) h'(r) h''(r) r^2-320 e^{2
	H(r)} q_2 \mathit{k} f(r)^2 h'(r) H'(r) h''(r) r^2\nonumber\\&&-64 e^{2 H(r)} q_2 \mathit{k} f(r)^2 h'(r)^2 H''(r)
r^2-192 e^{2 H(r)} q_2 \mathit{k} f(r) f'(r) h'(r)^2 r\nonumber\\&&+384 e^{2 H(r)} q_2 \mathit{k} f(r)^2 h'(r)^2 H'(r)
r+128 e^{2 H(r)} q_2 \mathit{k} f(r)^2 h'(r) h''(r) r\nonumber\\&&-144 e^{2 H(r)} q_2 \mathit{k} f(r)^2 h'(r)^2=0.
\end{eqnarray}

Non-zero component of Eq.(\ref{EOM-Maxwell}) is as follows,\\
\begin{eqnarray}
&&-32 q_2 f(r)^3 h'(r) H'(r)^5 r^5+32 q_2 f(r)^2 f'(r) h'(r) H'(r)^4 r^5+24 q_2 f(r) f'(r)^2 h'(r)
H'(r)^3 r^5\nonumber\\&&+40 q_2 f(r) h'(r) H'(r) f''(r)^2 r^5-160 q_2 f(r)^2 f'(r) h'(r) H''(r)^2 r^5\nonumber\\&&+96 q_2
f(r)^3 h'(r) H'(r) H''(r)^2 r^5-32 q_2 f(r)^3 h''(r) H''(r)^2 r^5+2 q_1 f(r) h'(r) H'(r) r^5\nonumber\\&&+64
q_2 f(r)^2 h'(r) H'(r)^3 f''(r) r^5-128 q_2 f(r) f'(r) h'(r) H'(r)^2 f''(r) r^5\nonumber\\&&-32 q_2 f(r)^3
H'(r)^4 h''(r) r^5+96 q_2 f(r)^2 f'(r) H'(r)^3 h''(r) r^5-72 q_2 f(r) f'(r)^2 H'(r)^2 h''(r) r^5\nonumber\\&&-8
q_2 f(r) f''(r)^2 h''(r) r^5+2 q_1 f(r) h''(r) r^5-32 q_2 f(r)^2 H'(r)^2 f''(r) h''(r) r^5\nonumber\\&&+48
q_2 f(r) f'(r) H'(r) f''(r) h''(r) r^5-64 q_2 f(r)^3 h'(r) H'(r)^3 H''(r) r^5\nonumber\\&&+320 q_2 f(r)^2 f'(r)
h'(r) H'(r)^2 H''(r) r^5-240 q_2 f(r) f'(r)^2 h'(r) H'(r) H''(r) r^5\nonumber\\&&+80 q_2 f(r) f'(r) h'(r) f''(r)
H''(r) r^5-128 q_2 f(r)^2 h'(r) H'(r) f''(r) H''(r) r^5\nonumber\\&&+64 q_2 f(r)^3 H'(r)^2 h''(r) H''(r) r^5-96
q_2 f(r)^2 f'(r) H'(r) h''(r) H''(r) r^5\nonumber\\&&+32 q_2 f(r)^2 f''(r) h''(r) H''(r) r^5-32 q_2 f(r)^2
h'(r) H'(r)^2 f^{(3)}(r) r^5\nonumber\\&&+48 q_2 f(r) f'(r) h'(r) H'(r) f^{(3)}(r) r^5-16 q_2 f(r) h'(r) f''(r)
f^{(3)}(r) r^5\nonumber\\&&+32 q_2 f(r)^2 h'(r) H''(r) f^{(3)}(r) r^5+64 q_2 f(r)^3 h'(r) H'(r)^2 H^{(3)}(r) r^5\nonumber\\&&-96
q_2 f(r)^2 f'(r) h'(r) H'(r) H^{(3)}(r) r^5+32 q_2 f(r)^2 h'(r) f''(r) H^{(3)}(r) r^5\nonumber\\&&-64 q_2 f(r)^3
h'(r) H''(r) H^{(3)}(r) r^5+64 q_2 f(r)^3 h'(r) H'(r)^4 r^4\nonumber\\&&+128 q_2 f(r)^2 f'(r) h'(r) H'(r)^3 r^4-272
q_2 f(r) f'(r)^2 h'(r) H'(r)^2 r^4\nonumber\\&&-80 q_2 f(r) h'(r) f''(r)^2 r^4-192 q_2 f(r)^3 h'(r) H''(r)^2
r^4+4 q_1 f(r) h'(r) r^4\nonumber\\&&-320 q_2 f(r)^2 h'(r) H'(r)^2 f''(r) r^4+480 q_2 f(r) f'(r) h'(r) H'(r)
f''(r) r^4\nonumber\\&&+128 q_2 f(r)^3 H'(r)^3 h''(r) r^4-320 q_2 f(r)^2 f'(r) H'(r)^2 h''(r) r^4\nonumber\\&&+192 q_2 f(r)
f'(r)^2 H'(r) h''(r) r^4-64 q_2 f(r) f'(r) f''(r) h''(r) r^4\nonumber\\&&+64 q_2 f(r)^2 H'(r) f''(r) h''(r) r^4+384
q_2 f(r)^3 h'(r) H'(r)^2 H''(r) r^4\nonumber\\&&+320 q_2 f(r) f'(r)^2 h'(r) H''(r) r^4-960 q_2 f(r)^2 f'(r)
h'(r) H'(r) H''(r) r^4\nonumber\\&&+256 q_2 f(r)^2 h'(r) f''(r) H''(r) r^4+128 q_2 f(r)^2 f'(r) h''(r) H''(r)
r^4\nonumber\\&&-128 q_2 f(r)^3 H'(r) h''(r) H''(r) r^4-64 q_2 f(r) f'(r) h'(r) f^{(3)}(r) r^4\nonumber\\&&+64 q_2 f(r)^2
h'(r) H'(r) f^{(3)}(r) r^4+128 q_2 f(r)^2 f'(r) h'(r) H^{(3)}(r) r^4\nonumber\\&&-128 q_2 f(r)^3 h'(r) H'(r)
H^{(3)}(r) r^4-64 q_2 f(r)^3 h'(r) H'(r)^3 r^3\nonumber\\&&-352 q_2 f(r)^2 f'(r) h'(r) H'(r)^2 r^3+416 q_2 f(r)
f'(r)^2 h'(r) H'(r) r^3\nonumber\\&&-352 q_2 f(r) f'(r) h'(r) f''(r) r^3+384 q_2 f(r)^2 h'(r) H'(r) f''(r) r^3\nonumber\\&&-128
q_2 f(r) f'(r)^2 h''(r) r^3-192 q_2 f(r)^3 H'(r)^2 h''(r) r^3+352 q_2 f(r)^2 f'(r) H'(r) h''(r)
r^3\nonumber\\&&-32 q_2 f(r)^2 f''(r) h''(r) r^3+608 q_2 f(r)^2 f'(r) h'(r) H''(r) r^3-448 q_2 f(r)^3 h'(r)
H'(r) H''(r) r^3\nonumber\\&&+64 q_2 f(r)^3 h''(r) H''(r) r^3-32 q_2 f(r)^2 h'(r) f^{(3)}(r) r^3+64 q_2 f(r)^3
h'(r) H^{(3)}(r) r^3\nonumber\\&&+128 q_2 f(r)^3 h'(r) H'(r)^2 r^2-128 q_2 f(r) f'(r)^2 h'(r) r^2+128 q_2 f(r)^2
f'(r) h'(r) H'(r) r^2\nonumber\\&&-128 q_2 f(r)^2 h'(r) f''(r) r^2-128 q_2 f(r)^2 f'(r) h''(r) r^2+128 q_2
f(r)^3 H'(r) h''(r) r^2\nonumber\\&&+128 q_2 f(r)^3 h'(r) H''(r) r^2+64 q_2 f(r)^2 f'(r) h'(r) r\nonumber\\&&-160 q_2 f(r)^3
h'(r) H'(r) r-32 q_2 f(r)^3 h''(r) r+64 q_2 f(r)^3 h'(r)=0,
\end{eqnarray}

We write the $tt$-component of Einstein's equation, Eq. (\ref{tt-comp}), up to first order as, 
\begin{eqnarray}\label{tt1}
&&-12 \mathit{k} r^4 f_0''(r)^2 h_0'(r)^2+16 \mathit{k} r^4 f_0'(r) f_0''(r) h_0'(r)
h_0''(r)+64 \mathit{k} r^4 f_0(r) f_0^{(3)}(r) h_0'(r) h_0''(r)\nonumber\\&&+32 \mathit{k} r^4 f_0(r)
h_0^{(3)}(r) f_0''(r) h_0'(r)+128 \mathit{k} r^3 f_0(r) f_0'(r) h_0''(r)^2-48 \mathit{k}
r^3 f_0'(r) f_0''(r) h_0'(r)^2\nonumber\\&&+64 \mathit{k} r^3 f_0'(r)^2 h_0'(r) h_0''(r)+320
\mathit{k} r^3 f_0(r) f_0''(r) h_0'(r) h_0''(r)+160 \mathit{k} r^2 f_0(r) f_0'(r)
h_0'(r) h_0''(r)\nonumber\\&&+8 \mathit{k} r^4 f_0^{(3)}(r) f_0'(r) h_0'(r)^2+16 \mathit{k} r^4
f_0(r) f_0^{(4)}(r) h_0'(r)^2+96 \mathit{k} r^3 f_0(r) f_0^{(3)}(r) h_0'(r)^2\nonumber\\&&+128
\mathit{k} r^3 f_0(r) h_0^{(3)}(r) f_0'(r) h_0'(r)-144 \mathit{k} r^2 f_0'(r)^2
h_0'(r)^2-192 \mathit{k} r f_0(r) f_0'(r) h_0'(r)^2\nonumber\\&&+32 \mathit{k} r^4 f_0(r) f_0''(r)
h_0''(r)^2-128 \mathit{k} r f_0(r)^2 h_0'(r) h_0''(r)+64 \mathit{k} r^2 f_0(r)^2
h_0^{(3)}(r) h_0'(r)\nonumber\\&&+48 \mathit{k} f_0(r)^2 h_0'(r)^2+64 \mathit{k} r^2 f_0(r)^2
h_0''(r)^2+2 r^3 f_1'(r)+2 r^2 f_1(r)+2 \mathit{k} q_1 r^4 h_0'(r) h_1'(r)\nonumber\\&&+2 \mathit{k}
q_1 r^4 H_1(r) h_0'(r)^2
\end{eqnarray}

We write the $rr$-component of Einstein's equation, Eq. (\ref{rr-comp}), up to first order as, 
\begin{eqnarray}\label{rr1}
&&-12 \mathit{k} r^4 f_0''(r)^2 h_0'(r)^2+16 \mathit{k} r^4 f_0'(r) f_0''(r) h_0'(r) h_0''(r)-48 \mathit{k} r^3
f_0'(r) f_0''(r) h_0'(r)^2\nonumber\\&&+64 \mathit{k} r^3 f_0'(r)^2 h_0'(r) h_0''(r)+64 \mathit{k} r^3 f_0(r)
f_0''(r) h_0'(r) h_0''(r)+96 \mathit{k} r^2 f_0(r) f_0''(r) h_0'(r)^2\nonumber\\&&+288 \mathit{k} r^2 f_0(r) f_0'(r)
h_0'(r) h_0''(r)+8 \mathit{k} r^4 f_0^{(3)}(r) f_0'(r) h_0'(r)^2+32 \mathit{k} r^3 f_0(r) f_0^{(3)}(r)
h_0'(r)^2\nonumber\\&&-144 \mathit{k} r^2 f_0'(r)^2 h_0'(r)^2-192 \mathit{k} r f_0(r) f_0'(r) h_0'(r)^2+128 \mathit{k} r
f_0(r)^2 h_0'(r) h_0''(r)\nonumber\\&&-144 \mathit{k} f_0(r)^2 h_0'(r)^2-4 r^3 f_0(r) H_1'(r)+2 r^3 f_1'(r)+2 r^2
f_1(r)\nonumber\\&&+2 \mathit{k} q_1 r^4 h_0'(r) h_1'(r)+2 \mathit{k} q_1 r^4 H_1(r) h_0'(r)^2
\end{eqnarray}
We impose the condition that Eq. (\ref{tt1}) and Eq. (\ref{rr1}) must be equivalent up to the first order of $q_2$. 
\begin{eqnarray}
	&&-64 \mathit{k} r^4 f_0(r) f_0^{(3)}(r) h_0'(r) h_0''(r)-32 \mathit{k} r^4 f_0(r)
h_0^{(3)}(r) f_0''(r) h_0'(r)-128 \mathit{k} r^3 f_0(r) f_0'(r) h_0''(r)^2\nonumber\\&&-256 \mathit{k}
r^3 f_0(r) f_0''(r) h_0'(r) h_0''(r)+96 \mathit{k} r^2 f_0(r) f_0''(r)
h_0'(r)^2+128 \mathit{k} r^2 f_0(r) f_0'(r) h_0'(r) h_0''(r)\nonumber\\&&-16 \mathit{k} r^4 f_0(r)
f_0^{(4)}(r) h_0'(r)^2-64 \mathit{k} r^3 f_0(r) f_0^{(3)}(r) h_0'(r)^2-128 \mathit{k} r^3
f_0(r) h_0^{(3)}(r) f_0'(r) h_0'(r)\nonumber\\&&-32 \mathit{k} r^4 f_0(r) f_0''(r)
h_0''(r)^2+256 \mathit{k} r f_0(r)^2 h_0'(r) h_0''(r)-64 \mathit{k} r^2 f_0(r)^2
h_0^{(3)}(r) h_0'(r)\nonumber\\&&-192 \mathit{k} f_0(r)^2 h_0'(r)^2-64 \mathit{k} r^2 f_0(r)^2
h_0''(r)^2-4 r^3 f_0(r) H_1'(r)=0,
\end{eqnarray}

\section{Scrambling time}\label{scram}
To compute the scrambling time we need to find the butterfly velocity. In the presence of a particle with momentum $P$ the energy-momentum tensor due to backreaction would be \cite{roberts}:
\begin{align}
T_{UU}\sim P e^{\frac{2\pi t}{\beta}}\delta (U)\delta (x)-2K(x,t)\delta(U)T^0_{UV},
\end{align}
$T^0_{UV}$ is the energy-momentum without the shock wave, where the backreacted geometry in the Kruskal coordinates would be:
\begin{align}
ds^2=2M(UV)dUdV+N(UV)dx^idx^j-2M(UV)K(x,t)\delta(U)dU^2.
\end{align}
By plugging the above metric in the Einstein equation:
\begin{align}
(\nabla_i-\lambda^2)K\sim \frac{N(0)Pe^{\frac{2\pi t}{\beta}}\delta(x)}{M(0)},
\end{align}
where $\lambda^2=\frac{d}{2}\frac{N'(0)}{M(0)}$ is the effective mass. The solution of $K$ function would be:
\begin{align}
K(x,t)=\frac{e^{\frac{2\pi (t-t_*)}{\beta}-\lambda x}}{|x|^{\frac{d-1}{2}}}
\end{align}
Then, the scrambling time would be:
\begin{align}
v_B=\frac{2\pi}{\beta \lambda}
\end{align}
For our solution \eqref{metric} the butterfly velocity would be:
\begin{align}
v^2_B&\sim \frac{Te^{-H(R)}}{R}=\frac{f'(R)e^{-2H(R)}}{4\pi R}= \frac{e^{\frac{2q_2Q^2}{R^4}}}{4\pi R}f'(R)\\
&\sim 1-Q^2(1+q_2)
\end{align}
Finally, the butterfly velocity is given by (setting the horizon as unity):
\begin{align}
v_B\sim \sqrt{1-Q^2(1+q_2)}
\end{align}
Then, the scrambling time would be \cite{swingle}:
\begin{align}
&t_*(x)\approx \frac{1}{\lambda_L}\Big(\log S+\frac{x\lambda_L}{v_B}\Big),\label{st}\\
&S\sim x^2,
\end{align}
where the Lyapunov exponent is $\lambda_L=\frac{\beta}{2\pi}$ and $x$ is the size of the system. The butterfly velocity in our solution is small, then the dominant term in the scrambling time is $\frac{x}{v_B}$:
\begin{align}
t_*\sim \frac{1}{\sqrt{1-Q^2(1+q_2)}}.
\end{align}


\end{document}